\documentclass[12pt]{iopart}
\usepackage{doi}
\usepackage{upgreek}
\usepackage{cite}
\usepackage{iopams}
\usepackage{xcolor}
\newcommand{\dms}[1]{\hat{\uprho}_{B}(#1)}
\usepackage{graphicx}
\graphicspath{{./All_images/}}
\usepackage{caption}
\usepackage{subcaption}
\usepackage{circuitikz}
\usepackage{relsize}
\begin{document}

\title{Collisional charging of a transmon quantum battery}
\author{Nicolò Massa$^{1}$, Fabio Cavaliere$^{1,2}$ and  Dario Ferraro$^{1,2,*}$}
\address{%
$^{1}$ \quad  Dipartimento di Fisica, Università di Genova, Via Dodecaneso 33, 16146, Genova, Italy \\
$^{2}$ \quad CNR-SPIN, Via Dodecaneso 33, 16146, Genova, Italy}
\vspace{10pt}
\begin{indented}
\item[] June 18, 2025
\end{indented}
\begin{abstract}
Motivated by recent developments in the field of multilevel quantum batteries, we present the model of a quantum device for energy storage with anharmonic level spacing, based on a superconducting circuit in the transmon regime. It is charged via the sequential interaction with a collection of identical and independent ancillary two-level systems. By means of a numerical analysis we show that, in case these ancillas are coherent, this kind of quantum battery can achieve remarkable performances for what it concerns the control of the stored energy and its extraction in regimes of parameters within reach in nowadays quantum circuits.
\end{abstract}

%
%
%
%
%

\section{Introduction}
In the last decades, the possibility to control the puzzling features of quantum mechanics and exploit them to realize devices able to outperform their classical counterparts opened the way to the development of several promising quantum technologies~\cite{Koch22, Ezratty24}, among which quantum batteries (QBs) have recently emerged as a hot topic~\cite{Quach23,Campaioli}. Indeed, since the seminal paper by Alicki and Fannes~\cite{Alicki13}, numerous proposals appeared with the aim of realizing miniaturized devices taking advantage of genuine non-classical features to improve energy storage, charging power and work extraction. A large part of these ideas is based on collections of two-level systems (TLSs) directly coupled among them~\cite{Le19, Rossini19, Rossini20, Grazi24, Catalano24, Porta20, Grazi25, Lu25} or via photons trapped into a resonant cavity~\cite{Ferraro18, Eckhardt22, Quach22, Carrasco22, Dou22, Gemme23, Wang24, Yang24, Erdman24, Hymas25}. 

Quite recently, experimental investigations of QBs realized using superconducting circuits in the transmon regime~\cite{Koch, Krantz_2019} have been reported~\cite{Hu22, Gemme24}. This has triggered new studies devoted to exploit the multilevel structure of quantum systems as a way to achieve a greater and more efficient energy storage~\cite{Din_blockade}. Among multilevel quantum systems, the harmonic oscillator plays a leading role due to its universality. It is therefore natural that, also in the QB literature, proposals for implementable multilevel devices based on quantum harmonic oscillators, experimentally realized for example by employing quantum LC circuits~\cite{Blais21}, have been considered~\cite{Hovhannisyan20, Rodriguez23}. 

A particularly insightful way to describe the charging of these harmonic oscillator QBs is based on the so-called collisional models~\cite{qubit_env,Q_bit_decoherence, Ciccarello, Cusumano_CM,Campbell_2021_CM,Q_info_and_thermo_Rep_interactions, Rndm_coll_heat_engine}. Originally introduced as a way to treat quantum dissipative systems~\cite{Petruccione,Ciccarello}, they are based on the decomposition of the dissipative environment into a collection of elementary building blocks, usually indicated as ancillas, which interact sequentially with the quantum system of interest. In general, the irreversible interaction with the environment is non-Markovian and typically leads to thermalization and energy losses from the system~\cite{Petruccione,Weiss,Morrone,qubit_env,Q_bit_decoherence,Ciccarello}. However, the same collisional models can be also used as an active way to charge a QB via a proper control of the quantum state of the ancillas and of the duration of their interaction with the QB~\cite{Barra19, Carrega_2020, Seah, Landi_Battery_CM}.\\
Particularly interesting in this direction is the case where the QB is a resonant cavity~\cite{Shaghaghi_MM, Salvia, Shaghaghi_Entropy_MM, Rodriguez23b}. This configuration is reminiscent of the micromaser experiments carried out by S. Haroche and coworkers where quantum states of radiation were engineered and characterized via a controlled flow of Rydberg atoms~\cite{Haroche13}. For properly fine tuned values of the parameters, this charging protocol leads to a very stable energy storage due to the emergence of trapping chambers ~\cite{Slosser89}. Here, after a given number of collisions with the ancillas, the QB reaches a configuration where the energy cannot be further reduced or increased and remains stably stored. As emerges from these works, the key advantage presented by collisional charging with respect to more conventional charging protocols for QBs, is that they allow the transfer of a greater amount of energy to the battery. Indeed, each charger intervening in the charging process can contribute to the building of an highly excited state of the battery, a goal which is not often achievable through the coupling with a single charger.

In view of possible implementations in solid-state platforms it is relevant to investigate the same collisional charging protocol in state of the art quantum circuits such as the transmons~\cite{Koch, Dou23}. Here, the multilevel spectrum is characterized by anharmonicity, which is the essential ingredient to properly realize effective two-level systems, playing the role of qubits for quantum computing~\cite{Krantz_2019, Roth_2023}. In this direction, the present paper aims at characterizing the charging and energy extraction efficiency of a transmon QB charged by means of a collisional approach. Through numerical analysis, we will show that the charging realized by means of independent and coherent ancillas leads to a (slightly damped) oscillating behavior as a function of the number of collisions whose functional dependence on the battery-ancilla coupling and coherence within the ancillas can be characterized through best fits. For short enough duration of the collisions, the energy stored into the QB remains stably within the cosine potential well induced by the Josephson energy of the transmon~\cite{devoret_sc_qubits} and, in correspondence of the first maxima of the oscillations, it can be almost completely extracted as useful work. Here, both the period of the oscillations and the damping depends on the circuit-ancilla coupling and, quite unexpectedly, on the coherence of the ancillas and not on their energy. For longer interaction durations, despite the greater amount of stored energy, the system shows beats and instabilities which progressively leads to a reduction of the efficiency in the energy extraction. When the charging is realized by means of incoherent ancillas the stored energy shows a completely different behavior as a function of the number of collisions: the oscillations disappear and the value of the stored energy depends on the coupling and on the energy initially trapped into the ancillas. In addiction, the efficiency of the energy extraction only approaches $50\%$.  

The article is organized as follows. In section \ref{Model} we describe the model for a transmon QB. The relevant parameters characterizing the device, the details of the considered collisional charging protocol and the relevant figures of merit are also discussed. Section \ref{Results} reports the results achieved by numerically studying the transmon QB in presence of repeated interactions with coherent and incoherent ancillas, for different values of the circuit-ancilla coupling, initial state of the ancillas (assumed identical) and times of collision. Considerations about the actual experimental feasibility are also reported here. In Section \ref{Conclusions} we comment on the results and draw some conclusions. For completeness, Appendices A and B are devoted respectively to the discussion of the, experimentally very challenging, regime of short times of collision, and to the effects of the reduction of ancillary coherences.


\section{Model}
\label{Model}
\subsection{Quantum battery}
\begin{figure}[h!]
\centering
\begin{subfigure}[b]{0.445\textwidth}
    \centering
    \includegraphics[width=\textwidth]{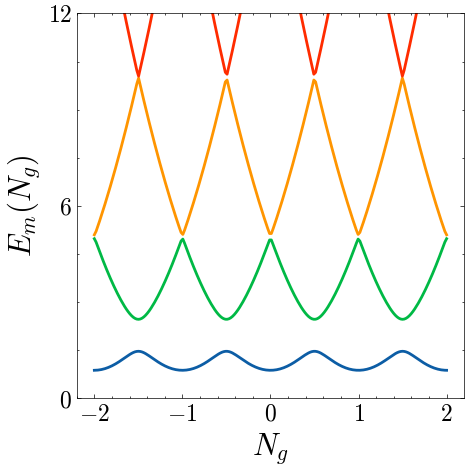}
    \caption{$\frac{E_J}{E_C}=1$}
\end{subfigure}   
\hspace{0.5cm}
\begin{subfigure}[b]{0.45\textwidth}
    \centering
    \includegraphics[width=\textwidth]{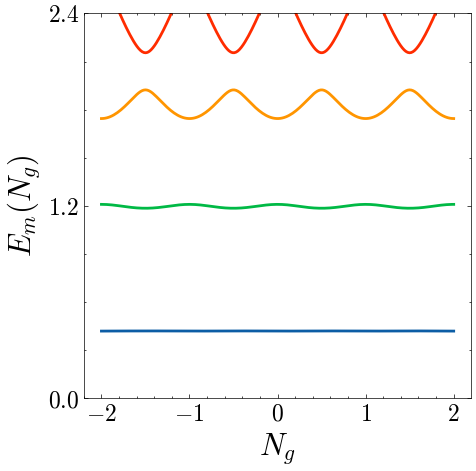}
    \caption{$\frac{E_J}{E_C}=10$}
\end{subfigure}
\vspace{0.5cm}
\begin{subfigure}[b]{0.45\textwidth}
    \centering
    \includegraphics[width=\textwidth]{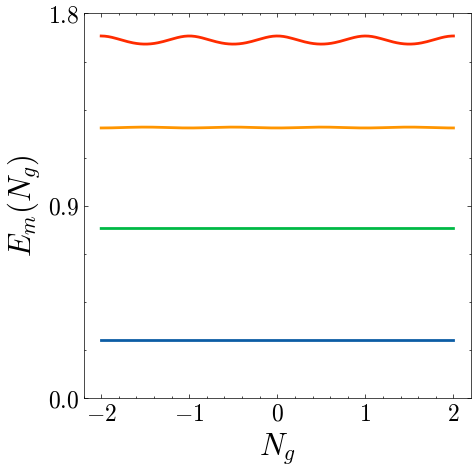}
    \caption{$\frac{E_J}{E_C}=25$}
\end{subfigure}   
\hspace{0.5cm}
\begin{subfigure}[b]{0.45\textwidth}
    \centering
    \includegraphics[width=\textwidth]{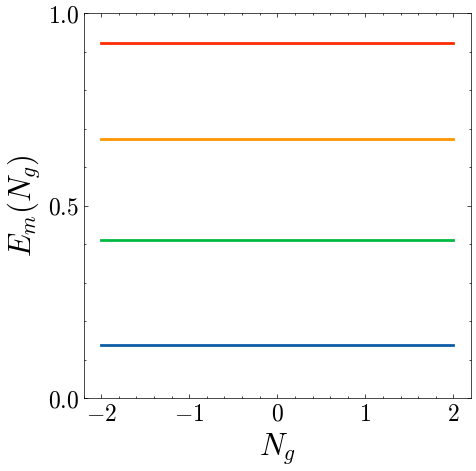}
    \caption{$\frac{E_J}{E_C}=100$}
\end{subfigure}  
 
\caption{Plot of the energy levels $E_m$ of the Hamiltonian in Eq.~(\ref{H_transmon}) (in units of $E_{J}$) for $m=0$ (blue), $1$ (green), $2$ (orange), $3$ (red) as functions of $N_g$ for different values of the ratio $E_J/E_C$.}
\label{fig:TR_LEVELS}
\end{figure}

\begin{figure}[h]
    \centering
    \includegraphics[width=0.5\textwidth]{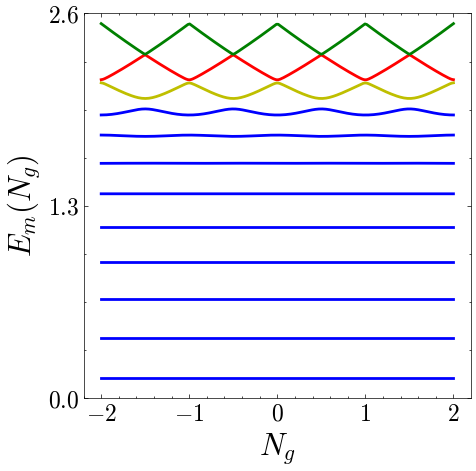}
    \caption{Plot of the energy levels $E_m$ of the Hamiltonian in Eq.~(\ref{H_transmon}) (in units of $E_{J}$) as functions of $N_g$ for $E_J/E_C=100$. States shown are both the trapped states $m=0,..,8$ (represented with blue curves) and the states outside the potential well $m=9$ (yellow), $m=10$ (red) and $m=11$ (green).}
    \label{fig:Spectrum}
\end{figure}

As a model for multilevel QB we consider a superconducting circuit in the transmon regime (see scheme in Fig.~\ref{fig:TRANSMON}). It is realized using a SQUID~\cite{Squid} shunted with a large capacitance $C_B$ and connected to a gate potential $V_g$ via a further capacitance $C_{g}$~\cite{Koch}.
\begin{figure}[h]
\centering
\begin{circuitikz}
    \ctikzset{capacitors/scale=0.7}
    \draw
    (1.5,0) -- (3.5,0)
    (1.5,2) -- (3.5,2)
    (1.5,0) to[barrier, l^= 
 $C_J$,name=Ej] (1.5,2)
    (3.5,0) to[barrier] (3.5,2)
    (2.5,2) -- (2.5, 2.5)
    (2.5,2.5) -- (5.5,2.5)

    (5.5,2.5) to[C, l^=$C_B$,name=Cb] (5.5,-0.5)
    (2.5,-0.5) -- (5.5, -0.5)
    (2.5,0) -- (2.5,-0.5)
    (4,3) to[C, l^=$C_g$,name=Cg] (-3,3)
    (4,3) -- (4,2.5)
    (4,-0.5) -- (4, -1)
    (-3,-1) -- (4,-1)
    (-3,3) to[battery1, l^=$V_g$,name=Vg] (-3,-1)
    (0.5,-1) -- (0.5,-1.2)

    (0,-1.2) -- (1,-1.2)
    (0.1,-1.3) -- (0.9,-1.3)
    (0.2,-1.4) -- (0.8,-1.4)
    (0.4,-1.5) -- (0.6,-1.5)
    ;
\end{circuitikz}
\vspace*{0.3cm}
\caption{Scheme of a transmon circuit. It is composed by a SQUID (realized by two Josephson junctions in parallel) with an intrinsic capacitance $C_{J}$. $C_B$ is a large external capacitance connected in parallel with the SQUID while $C_g$ indicates the gate capacitance and $V_g$ the gate voltage respectively.}
\label{fig:TRANSMON}
\end{figure}
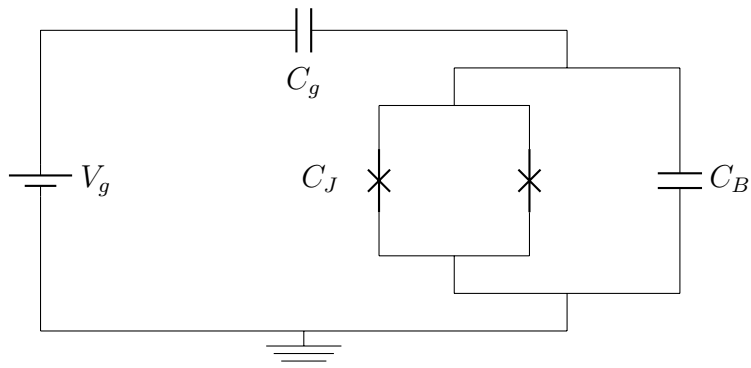\\
The Hamiltonian associated to this circuit can be written as (from now on we will assume $\hbar=1$) \cite{Koch,Blais21,devoret_sc_qubits}
\begin{equation}
    \hat{H}_{B} = 4E_C(\hat{N}-N_g)^2-E_J \cos(\hat{\varphi})
    \label{H_transmon}
\end{equation}
where $\hat{N}$ is the number operator of Cooper pairs transferred between the two superconducting islands forming the SQUID and $\hat{\varphi}$ is its conjugate variable, representing the superconducting phase difference across the SQUID. Together, they satisfy the commutation relation $\left[ \hat{\varphi}, \hat{N} \right]=i$. The parameter $E_C$, which represents the charging energy of the circuit, is given by $E_C = e^2/2C_{tot}$ with $e$ the electron charge and $C_{tot}=C_J+C_B+C_g$. On the other hand, $E_J$ is the Josephson energy associated with the tunneling of Cooper pairs~\cite{Koch}.
The parameter $N_g=C_{g}V_{g}/(2e)$ is an offset controlled by changing the gate potential $V_g$ and its role will be clarified later.\\
Since the cosine potential well in Eq.~(\ref{H_transmon}) is finite, $\hat{H}_B$ shows a discrete set of anharmonic bound levels within the well and a continuum of energy levels outside of it~\cite{Pietik_inen_2019}.
For generic values of the parameters, the Schr\"odinger equation for $\hat{H}_B$ can be solved, working in the basis of eigenstates of $\hat{\varphi}$, in terms of Mathieu's special functions~\cite{Koch}. This results in a set of anharmonic energy levels $\{E_m\}$ (with $m \in \mathbb{N}$) depending on $N_g,E_C,E_J$~\cite{Koch, 
Krantz_2019}.\\
The so-called transmon regime~\cite{Koch, 
Krantz_2019} is achieved when $E_J/E_C \gg 1$, a condition obtained in practice by using a large shunting capacitance $C_{B}$ to strongly reduce the charging energy of the circuit.\\
Figure~\ref{fig:TR_LEVELS} shows the behavior of the energy eigenvalues $E_m$ as a function of $N_{g}$ and for different values of $E_J/E_C$. In the regime $E_J\simeq E_C$ the energy levels are strongly dependent on $N_g$, so that the role of the gate voltage is to tune this parameter to the optimal working point of the circuit which, however, is still sensitive to gate fluctuations~\cite{Koch, devoret_sc_qubits}. Remarkably we observe that, by approaching the transmon regime (see Figs.~\ref{fig:TR_LEVELS} (a)-(d)), the dependence on $N_{g}$ of the low-energy levels progressively disappears, making the circuit almost immune to charge fluctuations.\\
Indeed, deep in the transmon regime, $\hat{H}_B$ can be approximated as a Duffing oscillator~\cite{Pietik_inen_2019,Duffing_Peano_2006,Duffing_PhysRevLett.99.137001,Duffing_Vierheilig} by Taylor expanding the cosine potential, obtaining
\begin{equation}
    \hat{H}_B = 4 E_C(\hat{N}-N_g)^2-E_J+\frac{E_J}{2}\hat{\varphi}^2-\frac{E_J}{24}\hat{\varphi}^4+o(E_J\hat{\varphi}^6).
\end{equation}
The justification of this expansion is that, if $E_C\ll E_J$, the height of the well is much greater than the kinetic part of Eq.~(\ref{H_transmon}), so that the system remains near to $\varphi=0$, the minimum of the well. The effectiveness of this expansion clearly emerges by writing $\hat{N}$ and $\hat{\varphi}$ in terms of the ladder operators $\hat{b}$ and $\hat{b}^\dagger$ of a quantum harmonic oscillator, namely
\begin{eqnarray}
    \hat{N}-&N_g =& \frac{i}{4}\sqrt{\frac{\omega_p}{E_C}}\left(\hat{b}^\dagger-\hat{b}\right)\\
    &\hat{\varphi} =& 2\sqrt{\frac{E_C}{\omega_p}}\left(\hat{b}^\dagger+\hat{b}\right)
\end{eqnarray}
where $\omega_p=\sqrt{8 E_J E_C}$ is the Josephson plasma frequency of the transmon.
These relations guarantee that the canonical commutation relation between $\hat{N}$ and $\hat{\varphi}$ is satisfied if $\left[\hat{b},\hat{b}^\dagger\right]=\mathbb{I}$.
\\
Writing the resulting Hamiltonian in units of the plasma frequency one obtains
\begin{equation}
    \frac{\hat{H}_B}{\omega_p} = -\frac{E_J}{\omega_p}+\hat{b}^\dagger\hat{b}+\frac{\mathbb{I}}{2}-\frac{E_C}{12\omega_p}\left(\hat{b}^\dagger+\hat{b}\right)^4+o\left(\frac{E_C^2}{\omega_p^2}\right).
    \label{eq:Perturbative_hamilt}
\end{equation}
This helps identifying as a perturbative parameter the ratio $E_C/\omega_p =\sqrt{E_C/8E_J}\ll 1$, so that Eq~(\ref{eq:Perturbative_hamilt}) can be treated perturbatively to lowest order in $E_C/\omega_p$.
Such an approach leads to the following gate-independent expression for the low-energy levels in units of $\omega_p$~\cite{Koch}
\begin{equation}
    \frac{E_m}{\omega_p} \approx -\frac{E_J}{\omega_p}+m+\frac{1}{2}-\frac{E_C}{4\omega_p}(2m^2+2m+1)+o\left(\frac{E_C^2}{\omega_p^2}\right).
    \label{eq:perturbed_en_levels}
\end{equation}
As shown in Fig.~\ref{fig:Spectrum}, the relation derived above is a good approximation for the transmon energy levels within the well (blue curves), whereas fluctuations with respect to $N_g$ become relevant only outside of it.\\
\\
In the regime considered above, it is possible to estimate the relative anharmonicity $\alpha_r$ of the spectrum, defined as  
\begin{equation}
    \alpha_r(m) \equiv \frac{|\Delta E_{m+1}-\Delta E_{m}|}{\Delta E_0}
\end{equation}
with $\Delta E_{m}=E_{m+1}-E_{m}$. Making use of Eq.~(\ref{eq:perturbed_en_levels})
one obtains
\begin{equation}
     \alpha_r = \frac{E_C}{\omega_p-E_C} 
\end{equation}
which, since in the transmon limit $\omega_p \gg E_C$, reduces to
\begin{equation}
    \alpha_r \approx \sqrt{\frac{E_C}{8E_J}},
    \label{eq:rel_anarm}
\end{equation}
which is the same result obtained for the perturbative parameter in the transmon regime.
The same ratio appearing in Eq.~(\ref{eq:rel_anarm}) determines, at least approximately, the number $N_b$ of bound states in the cosine well of the transmon, given by~\cite{Pietik_inen_2019}
\begin{equation}
    N_b \approx \sqrt{\frac{E_J}{E_C}}\propto \frac{1}{\alpha_r}.
    \label{bound_states}
\end{equation}
These features are the main reasons behind our choice of the transmon superconducting circuit as the QB: by increasing $E_J/E_C$ the number of bound energy levels increases, leading to a multilevel structure that is both anharmonic and charge-insensitive.

\subsection{Charging protocol}
The QB is charged through the sequential interaction with a collection of TLSs, a process which fits in the framework of collisional models~\cite{Ciccarello}. According to this, we will then refer to the individual collisional units both as ancillas and as chargers, whereas we will indicate the single QB-charger interaction as a "collision".\\
From now on, we will denote by $\hat{\uprho}_{B}(n)$ the QB density matrix after $n$ steps of QB-charger interaction, namely $\dms{n}\equiv\dms{\tau_1+\dots+\tau_n}$ with $\tau_i$ the duration of each interaction, and with $\{\hat{\eta}_n\}_{n\geq1}$ the set of initial density matrices that describe the chargers.\\
The general dynamics of a system undergoing many collisions can be quite involved. In fact, interactions and correlations among the ancillas may lead to memory effects, so that the density matrix of the system after $n$ collisions
depends, in general, on all the previous steps of the sequence of interactions. Since in a collisional charging process ancillas are quantum systems which need to be properly engineered, this work has been carried out assuming the possibility of controlling them in such a way that they are non-interacting and initially uncorrelated (see Sec.~\ref{sec:Exp_feas} for further details). These conditions, together with the assumption of chargers which individually interact with the QB, guarantee a Markovian dynamics \cite{Ciccarello}, which means that the state of the system after the $n$-th interaction depends only on the state of the $n$-th charger and on the state of the QB at the end of the previous step ($\hat{\uprho}_{B}(n-1)$).\\
For simplicity, we also assumed a homogeneous collision model, with identical chargers (each charger prepared in the same quantum state) interacting with the QB through a formally identical coupling Hamiltonian and an identical interaction duration $\tau$.\\
In order to properly describe the collisional dynamics, one can start with the interaction between the system and a single charger. The Hamiltonian encoding the interaction between the transmon and the $n$-th charger consists of three terms
\begin{equation}
\hat{H}_{Bn}^{(t)}=\hat{H}_{B}+\hat{H}_n+\hat{V}_{Bn}^{(t)}
\end{equation}
where $\hat{H}_B$ and $\hat{H}_n$ respectively represent the free Hamiltonian of the QB, given by Eq.~(\ref{H_transmon}), and the free Hamiltonian of the $n$--th TLS charger, of the form
\begin{equation}
    \hat{H}_n = \frac{\Delta}{2}\hat{\sigma}^{z}_{n}
\end{equation}
with $\hat{\sigma}^{z}_{n}$ the conventional Pauli matrix along the $z$ direction and $\Delta$ the level spacing of the qubit. Its eigenstates will be denoted by $\{|0\rangle_n, |1\rangle_n$\}.\\
As for the coupling between the QB and the $n$--th charger, the following time-dependent interaction Hamiltonian is considered
\begin{equation}
    \hat{V}_{Bn}^{(t)} =g\hat{N}(\hat{\sigma}^{+}_{n}+\hat{\sigma}^{-}_{n})\theta \left(t-(n-1)\tau\right)\theta \left(n\tau-t\right)
\end{equation}
where $\hat{\sigma}^{+}_{n} = |1\rangle \langle 0 |_n$ and $\hat{\sigma}^{-}_{n} = |0 \rangle \langle 1|_n$. Thus, the coupling between the battery and the $n$--th charger is turned on at $t_{n-1}=(n-1)\tau$ (first collision labeled with $n=1$) and switched off at $t_n=n\tau$, so that the duration of each interaction is always $\tau$. The coupling strength $g$ is the same for each step considered, as required by the homogeneity assumption. As we will see when discussing the experimental feasibility of our battery, such kind of interaction Hamiltonian can be realized by capacitively coupling
two superconducting circuits in the transmon regime~\cite{Krantz_2019}, with the one playing the role of the ancilla characterized by a greater anharmonicity in such a way to be properly described as a TLS.\\
The entire work has been carried out neglecting thermal effects, due to the cryogenic working temperatures of the superconducting circuits forming the transmon battery, which usually are of the order of $T\approx10mK$~\cite{Buluta_2011}.\\
We assume that the first collision occurs at $t=0$, when the battery is prepared in its ground state and each ancilla is prepared in a given superposition of ground ($|0\rangle_{n}$) and excited state ($|1\rangle_{n}$), namely 
\begin{equation}
    \hat{\eta}_n = (1-q)|1\rangle \langle 1|_n+q|0\rangle \langle 0|_n+c\sqrt{q(1-q)}\bigl(|1 \rangle \langle 0|_n+|0 \rangle \langle 1|_n\bigr),
    \label{state_ancillas}
\end{equation}
in order to capture effects due to quantum coherences. In fact, as pointed out in~\cite{Seah}, the presence of quantum coherences in the collisional charging of an harmonic quantum battery can lead to noteworthy advantages in storage performances. The aim of our analysis is to investigate the possibility of taking advantages from these quantum coherences also in presence of anharmonicity.
Here, the real parameter $c$ has been introduced to distinguish between a perfectly coherent charging protocol ($c=1$) and an incoherent one $(c=0)$. At fixed $c$, the parameter $q$ can be used to control both the coherences of the charger and its populations.\\
The charging protocol then proceeds through a sequence of $N$ collisions, the $n$--th of which goes as follows: 
\begin{itemize}
    \item at the time $t_{n-1}=(n-1)\tau$, the QB is coupled with the $n$--th TLS charger, so that the resulting composite system undergoes a unitary evolution described by the Hamiltonian $\hat{H}_{Bn}^{(t)}$;
   \item at a time $t_{n}=n\tau$, the coupling is switched off;
   \item this procedure is iteratively repeated by coupling the QB with subsequent chargers, until one obtains the desired value of a relevant figure of merit (see below).
\end{itemize}
Starting from the following initial state for the whole system (QB + $N$ chargers) 
\begin{equation}
    \hat{\uprho}(0) = \hat{\uprho}_{B}(0)\otimes\hat{\eta}_1 \dots \otimes\hat{\eta}_N,
\end{equation}
which satisfies the requirement of initially uncorrelated ancillas, the Markovian time evolution of the QB density matrix can be obtained as \cite{Ciccarello}
\begin{equation}
     \hat{\uprho}_{B}(n) = M_n[M_{n-1}[...M_1[\dms{0}]]]
\end{equation}
where we have defined the quantum map 
\begin{equation}
    M_n[\hat{\uprho}] = \mathrm{Tr}_n\bigl\{
    \hat{U}_n \hat{\uprho} \otimes \hat{\eta}_n {U}^\dagger_n
    \bigr\},
\end{equation}
with $\hat{U}_n$ denoting the unitary evolution operator for the QB$+n$--th charger composite system. The role of this map in the dynamics of the system is to connect the QB density matrix at two subsequent steps. As usually done, for sake of simplicity, in the context of collision models, our calculations will be performed in the interaction picture. 
Since an analytical expression for the quantum map is out of reach for our model, we have employed a numerical approach based on the QuTip toolbox~\cite{Johansson12} to solve the stepwise evolution of $\dms{n}$. Within each system-ancilla interaction, the time evolution of $\hat{\uprho}_B$ has been determined
numerically solving the Von Neumann equation of the system-ancilla global density matrix and subsequently tracing over the ancillary degrees of freedom.

\subsection{Figures of merit}
The first figure of merit needed to characterize the performances of the QB is the energy stored as a function of time (here the number of collisions), defined as~\cite{Ferraro18, Andolina18}
\begin{equation}
    \Delta E(n) \equiv E(n)-E(0) = \mathrm{Tr}_{B}\{\hat{H}_{B}[\hat{\uprho}_{B}(n)-\hat{\uprho}_{B}(0)]\}.
\end{equation}
The extraction of such energy occurs via unitary operations acting on the battery~\cite{Binder15, Bhattacharjee21}. Such operations lead to a maximum amount of extracted energy $\mathcal{E}(n) \leq \Delta E(n)$, leaving the battery in a passive (empty) state~\cite{Passive_states}. The maximum energy that can be extracted via unitary operations is called ergotropy and is defined as~\cite{Allahverdian}
\begin{equation}
    \mathcal{E}(n) = \mathrm{Tr}_{B}\{\dms{n} \hat{H}_{B}\}-\min_{\hat{U} \in SU(d)}\mathrm{Tr}_{B}\{\hat{U} \dms{n} \hat{U}^\dagger \hat{H}_{B}\}.
    \label{Erg_def}
\end{equation}
Here, $\hat{U}$ represents a generic unitary discharging operation which is applied to the QB once the energy has been stored in it, in order to extract it as useful work. Notice that, denoting with $d$ the dimensionality of the battery Hilbert's space, in the second term a minimization in the space $SU(d)$ of the unitary operations acting on the $d$-dimensional quantum system is carried out.\\
Given the following spectral decompositions of the the battery Hamiltonian and state
\begin{eqnarray}
    \hat{H}_B &=&\sum_{j=1}^d E_j |E_j\rangle\langle E_j| \quad \mathrm{with} \quad  E_{j+1}\geq E_{j}\\
    \dms{n}&=&\sum_{j=1}^{d} p_j |p_j\rangle\langle p_j|\quad \mathrm{with} \quad p_{j}\geq p_{j+1},
\end{eqnarray}
the ergotropy can be evaluated without carrying out explicitly the $SU(d)$ minimization as \cite{Allahverdian}
\begin{equation}
    \mathcal{E}(n)= \sum_{j,k}^d p_j E_k \bigl(|\langle p_j|E_k\rangle|^2-\delta_{jk}\bigr).
\end{equation}
It is then useful to characterize the extraction process through the energy extraction efficiency of the QB, defined as~\cite{Hovhannisyan20, Barra22, Din_blockade} 
\begin{equation}
    \eta(n)=\frac{\mathcal{E}(n)}{E(n)}.
\end{equation}
This, together with the stored energy, is a key figure of merit which allows to identify optimal working conditions for the transmon QB: the higher $\eta$, the more available the stored energy actually is as useful work


\section{Results}
\label{Results}
In this Section we show numerical results, obtained via the QuTip toolbox~\cite{Johansson12}, concerning the collisional charging of a superconducting QB in the transmon regime. We will analyze the stored energy $\Delta E(n)$ and the efficiency $\eta(n)$ as functions of the collisional step $n$ in different regimes of the parameters.\\
Since, as stated above, the transmon QB bound states are insensitive to the value of $N_{g}$, we can safely set $N_g=0$ for all the following discussions without loss of generality. Moreover, the gap $\Delta$ of each charger will be assumed to be in resonance with the first energetic gap of the transmon spectrum, namely $\Delta =E_1-E_0$. A perfect fine-tuning of $\Delta$ is not needed, since slight deviations from this condition did not lead to major deviations from the results we are going to show. We expect, however, that a consistent reduction of the gap could cause disruptive effects, since it would bring the ancillas far from the resonance with the bound spectrum gaps of the transmon.\\
Recalling Eq.~\ref{state_ancillas}, two different situations will be considered: 
\begin{enumerate}
    \item[\emph{i)}] a \emph{coherent} charging protocol, where $c=1$ and $q \in (0,1)$, discussed in Sec.~\ref{sec:sec_coherent};
    \item[\emph{ii)}] an \emph{incoherent} charging protocol, where $c=0$, discussed in Sec.~\ref{sec:sec_incoherent}.
\end{enumerate} 
Both analysis will focus on coupling values in the regime $10^{-3} \lesssim g/\omega_p < 10^{-1}$, which fits the regime usually indicated as strong coupling, where the coupling is greater with respect to the system loss rates ~\cite{Kockum_2019_coupling}. \\
Concerning the interaction duration $\tau$ of each collision, considering as a reference time scale the inverse of the plasma frequency of the superconducting circuit $\tau_p=1/\omega_p$, we focused mainly on sequences of interactions with $\tau \geq \tau_p$. The regime with $\tau < \tau_p$ would require manipulations of the ancillary circuits faster than the scale $\tau_p$, whose typical values are of the order of a few nanoseconds
(see Sec.~\ref{sec:Exp_feas} for more details).
Given the technical difficulty in manipulating superconducting circuits at such short time scales, the regime $\tau < \tau_p$ appears less appealing in view of experimental realizations. Additional considerations about this regime are reported in Appendix A, whereas intermediate regimes of ancillary coherences will be discussed in Appendix B.


\subsection{Coherent charging}
\label{sec:sec_coherent}
When $c=1$ the off-diagonal entries of $\hat{\eta}_n$ are non-zero provided that $q \in (0,1)$, they are maximal at $q=0.5$ and symmetric with respect to this value in the same interval. This means that, by comparing two charging protocols with $q$ and $1-q$, we are making a comparison between situations where the chargers have the same coherences and exchanged populations. 
\subsubsection{Collision time $\tau=\tau_p$}
\label{sec:sec_coherent_weak}
\begin{figure}[h!]
\centering
\begin{subfigure}[b]{0.48\textwidth}
    \centering
    \includegraphics[width=\textwidth]{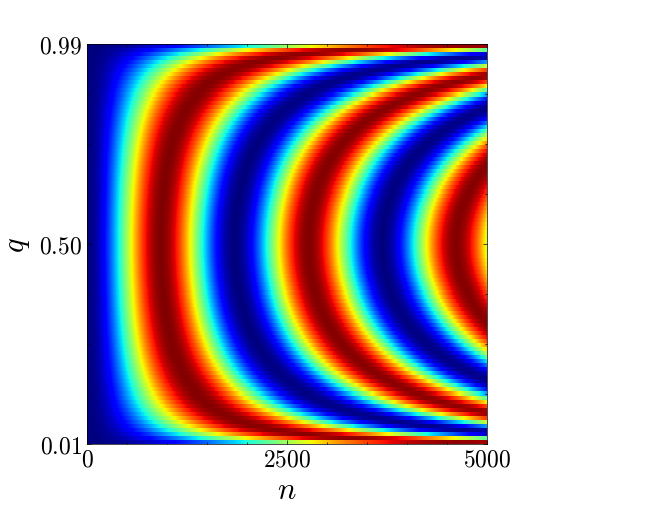}
    \caption{}
\end{subfigure}   
\hspace{0.1cm}
\begin{subfigure}[b]{0.48\textwidth}
    \centering
    \includegraphics[width=\textwidth]{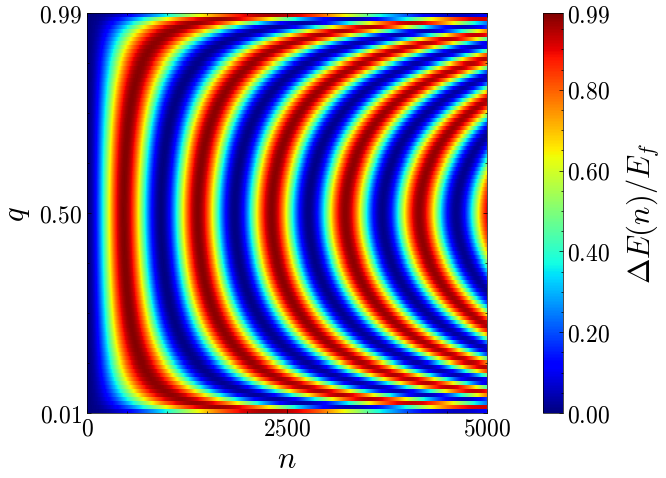}
    \caption{}
\end{subfigure} 
\caption{Density plots of the stored energy $\Delta E(n,q)$ (in units of $E_{f}=E_J-E_0$) as a function of the number of collisions $n$ and of the parameter $q$ for: (a) $g=4\times 10^{-3}\omega_p$, (b) $g=8\times 10^{-3}\omega_p$. Other parameters are: $E_J/E_C=100$, $\tau=\tau_p$ and $c=1$.}
\label{fig:TAUp_WEAK_COUPL_DENSITIES}
\end{figure}

\begin{figure}[h!]
\centering
\begin{subfigure}[b]{0.48\textwidth}
    \centering
    \includegraphics[width=\textwidth]{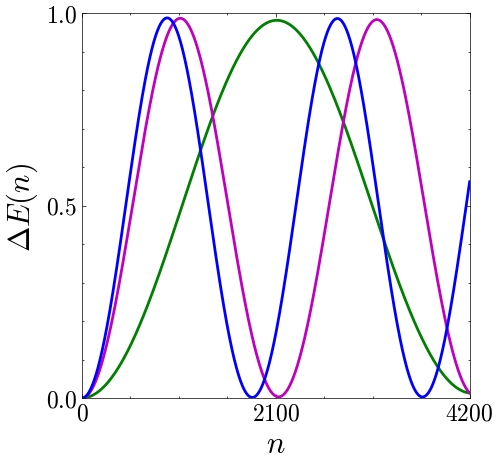}
    \caption{}
\end{subfigure}   
\hspace{0.1cm}
\begin{subfigure}[b]{0.48\textwidth}
    \centering
    \includegraphics[width=\textwidth]{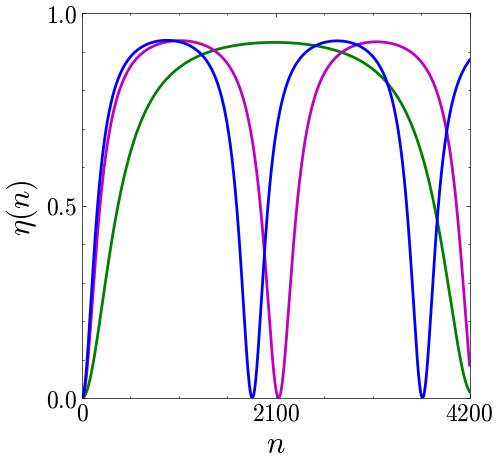}
    \caption{}
\end{subfigure} 
\caption{Stored energy $\Delta E(n)$ (in units of $E_f$) (a) and extraction efficiency $\eta(n)$ (b) as functions of the number of collisions $n$ for: $q=0.05$ (green line), $q=0.25$ (purple line), $q=0.5$ (blue line). Other parameters are: $E_J/E_C=100$, $\tau=\tau_p$, $g=4\times10^{-3}\omega_p$ and $c=1$.}
\label{fig:TAUp_WEAK_COUPL}
\end{figure}

\begin{figure}[h!]
\centering
\begin{subfigure}[b]{0.48\textwidth}
    \centering
    \includegraphics[width=\textwidth]{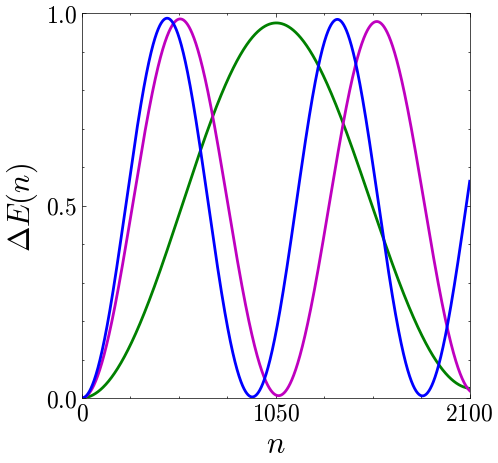}
    \caption{}
\end{subfigure}   
\hspace{0.1cm}
\begin{subfigure}[b]{0.48\textwidth}
    \centering
    \includegraphics[width=\textwidth]{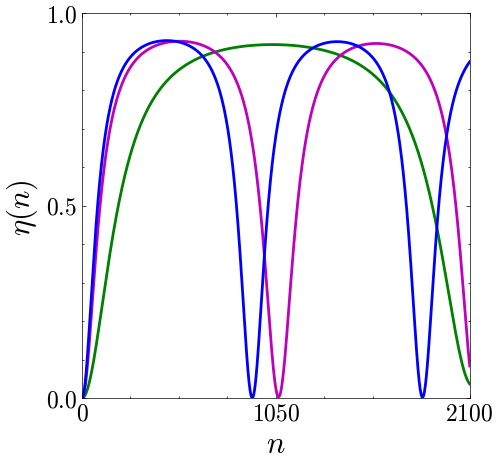}
    \caption{}
\end{subfigure} 
\caption{Stored energy $\Delta E(n)$ (in units of $E_f$) (a) and extraction efficiency $\eta(n)$ (b) as functions of the number of collisions $n$ for: $q=0.05$ (green line), $q=0.25$ (purple line), $q=0.5$ (blue line). Other parameters are: $E_J/E_C=100$, $\tau=\tau_p$, $g=8\times10^{-3}\omega_p$ and $c=1$.}
\label{fig:TAUp_WEAK_COUPL_008}
\end{figure}

For this regime, results are summarized in Fig.~\ref{fig:TAUp_WEAK_COUPL_DENSITIES}, which show the stored energy as a function of $n$ and $q$ for two different values of the coupling $g$.
For each of these, if $q$ is fixed, the QB oscillates between the ground state and a charged state where the energy approaches $E_{f} = E_J-E_0$, namely the top of the cosine potential well.
This guarantees that, provided one is able to stop the charging process at the right time, the QB can be driven from its ground state to the top of the potential well. \\
The stored energy is symmetric with respect to the $q=0.5$ axis, so that there is no difference between two charging protocols characterized by ancillary parameters $q_1$ and $q_2$ satisfying $q_1=1-q_2$.
This means that charging protocols with the same quantum coherences at the level of the chargers show the same evolution of the stored energy, regardless of their different populations (and energies). Thus, at least for what concerns values of $g$ in the lowest part of the coupling interval considered, the quantum coherences of the chargers seem to be the ultimate responsible for the QB dynamics. \\
Other interesting features emerging from Fig.~\ref{fig:TAUp_WEAK_COUPL_DENSITIES} concern the frequency $\Omega(g,q)$ of the energy oscillations. The more noticeable fact is that the frequency is higher at higher coupling. It is also symmetric with respect to $q=0.5$, where it reaches its maximum value, decreasing as $q$ moves towards $q=0$ or $q=1$.
These last points can be seen more easily in Figs.~\ref{fig:TAUp_WEAK_COUPL}(a) and ~\ref{fig:TAUp_WEAK_COUPL_008}(a) showing, for the values of the coupling introduced above, the stored energy as a function of $n$, for three different values of $q$. Clearly, Fig.~\ref{fig:TAUp_WEAK_COUPL_008}(a) shows the same number of oscillations of Fig.~\ref{fig:TAUp_WEAK_COUPL}(a), but within an halved period (doubled frequency). In addition, these figures show that the frequency decreases dramatically for $q\rightarrow0$, being maximal at $q=0.5$.\\
\\
Taking into account all these facts, and given the form of the coherent terms in the density matrix of the ancillas, a reasonable ansatz for the frequency $\Omega(g,q)$ seems to be
\begin{equation}
    \label{eq:ansatz_omega}
    \Omega(g,q)=\Omega^* \frac{g}{\omega_p} \sqrt{q(1-q)}
\end{equation}
with $\Omega^*$ being a proper scale factor.
We will further investigate this assumption in Sec.~\ref{sec:fits}.\\
Thus, by increasing the coupling, the frequency increases so that the charging is faster and is characterized by a higher power. This feature recalls the behavior of harmonic systems such as the Dicke QBs, where a greater QB-charger coupling leads to a faster charging ~\cite{Ferraro18,Ultrafast_Dicke_QB,Boosting_en_transfer}.\\

\begin{figure}[h!]
\centering
\begin{subfigure}[b]{0.48\textwidth}
    \centering
    \includegraphics[width=\textwidth]{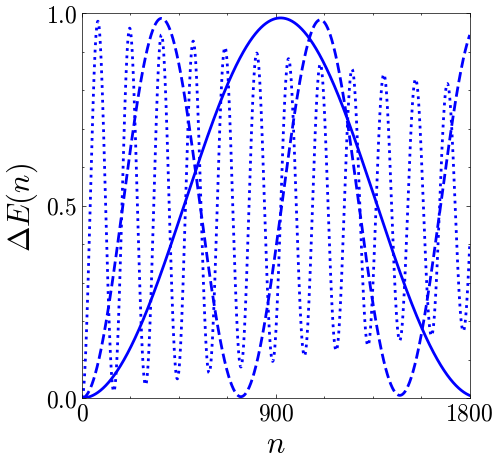}
    \caption{}
\end{subfigure}   
\hspace{0.1cm}
\begin{subfigure}[b]{0.48\textwidth}
    \centering
    \includegraphics[width=\textwidth]{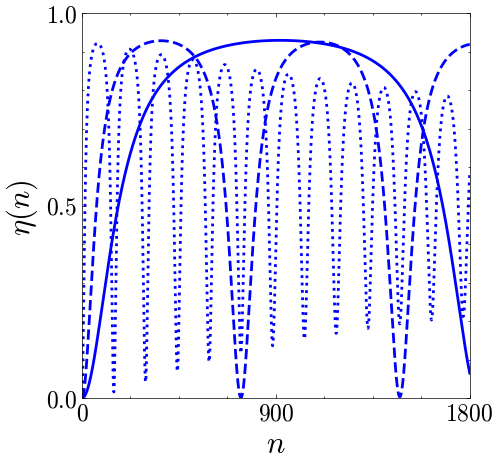}
    \caption{}
\end{subfigure} 
\caption{Stored energy $\Delta E(n)$ (in units of $E_{f}$) (a) and extraction efficiency $\eta(n)$ (b) as functions of the number of collisions $n$ for: $g/\omega_p=4\times10^{-3}$ (solid), $g/\omega_p=1\times10^{-2}$ (dashed), $g/\omega_p=5\times10^{-2}$ (dotted). Other parameters are: $E_J/E_C=100$, $\tau=\tau_p$, $q=0.5$ and $c=1$.}
\label{fig:TAUp_INCREASED_COUPL}
\end{figure}

\begin{figure}[h!]
\centering
\begin{subfigure}[b]{0.48\textwidth}
    \centering
    \includegraphics[width=\textwidth]{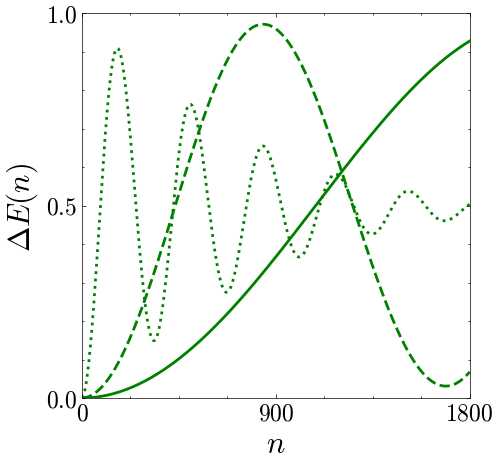}
    \caption{}
\end{subfigure}   
\hspace{0.1cm}
\begin{subfigure}[b]{0.48\textwidth}
    \centering
    \includegraphics[width=\textwidth]{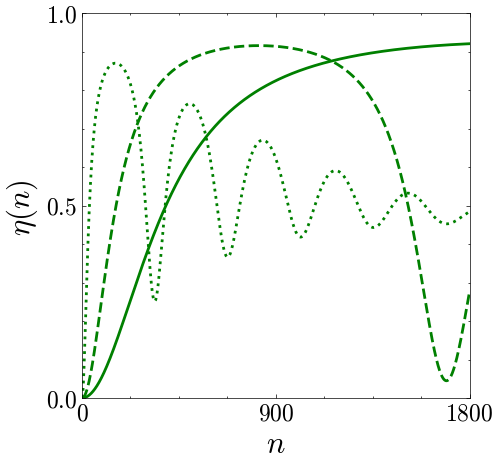}
    \caption{}
\end{subfigure} 
\caption{Stored energy $\Delta E(n)$ (a) and extraction efficiency $\eta(n)$ (b) as functions of the number of collisions $n$ for: $g/\omega_p=4\times10^{-3}$ (solid), $g/\omega_p=1\times10^{-2}$ (dashed), $g/\omega_p=5\times10^{-2}$ (dotted). Other parameters are: $E_J/E_C=100$, $\tau=\tau_p$, $q=0.05$ and $c=1$.}
    \label{fig:TAUp_INCREASED_COUPL_low_coher}
\end{figure}

Once discussed the behavior of the stored energy, the problem of efficient energy extraction can be addressed.
To this end, the results for the energy extraction efficiency $\eta(n)$ are shown in Figs.~\ref{fig:TAUp_WEAK_COUPL}(b) and~\ref{fig:TAUp_WEAK_COUPL_008}(b) for the two considered values of the coupling.\\
The oscillations performed by the efficiency reach a maximum value of $\eta \approx 0.9$ in correspondence of the energy maxima. We also notice that this quantity is flatter around its maxima than $\Delta E(n)$ . This is of great relevance, since it guarantees that a significant fraction of the stored energy can be extracted as useful work even without stopping the charging at its exact maximum.\\
Therefore, in this regime, the battery is able to store a large amount of energy (of the order of $E_f$):
\begin{enumerate}
    \item[\emph{i)}] in a short time $t_c=N_{max}\tau_p$ with $N_{max}$, the number of collisions needed to reach the first maximum, as little as $400-1000$, corresponding to a charging duration $t_c \approx 0.4-1$ $\mu$s (see Section ~\ref{sec:Exp_feas} for more details) ;
    \item[\emph{ii)}] with a very large extraction efficiency, reaching peaks with $\eta \approx 0.9$.\\
\end{enumerate}
An aspect which still needs to be investigated is how stable the discussed properties are with respect to variations of the coupling.
Results are shown for both the fastest and slowest charging cases considered before, namely $q=0.5$ (see Fig.~\ref{fig:TAUp_INCREASED_COUPL}) and $q=0.05$ (see Fig.~\ref{fig:TAUp_INCREASED_COUPL_low_coher}).
Focusing in particular on the stored energy, it is clear for both $q=0.5$ (see Fig.~\ref{fig:TAUp_INCREASED_COUPL} (a)) and $q=0.05$ (see Fig.~\ref{fig:TAUp_INCREASED_COUPL_low_coher} (a)) that increasing the coupling leads to faster oscillations (shorter period). \\
Moreover, one can note the emergence of a damping effect at high couplings (dotted lines) which is stronger for $q=0.05$ (low coherences of the ancillas), if compared with $q=0.5$ (high coherences of the ancillas), suggesting that the quantum coherences protect the battery from this damping.\\
All considered, two scenarios open: if one has a great control on the device, higher couplings can be useful to obtain a charged QB in short times (small number of collisions) avoiding the significant decrease of the stored energy; if, on the other hand, this fast control is not technically achievable, it could be convenient to work at weaker couplings where the effects of the damping become relevant only after a huge amount of collisions (see Figs.~\ref{fig:TAUp_WEAK_COUPL}(a) and ~\ref{fig:TAUp_WEAK_COUPL_008}(a)). The convenience of slower oscillations (weaker couplings) at long times is supported by the efficiency of the energy extraction, shown in Fig.~\ref{fig:TAUp_INCREASED_COUPL} (b) and Fig.~\ref{fig:TAUp_INCREASED_COUPL_low_coher} (b). Here, in fact, the stronger coupling exhibits the same efficiency as the weaker one only for few collisions, being characterized by a faster decay.

\subsubsection{Frequency and damping of the stored energy oscillations}
\label{sec:fits}
\begin{figure}[h!]
\centering
\begin{subfigure}[b]{0.48\textwidth}
    \centering
    \includegraphics[width=\textwidth]{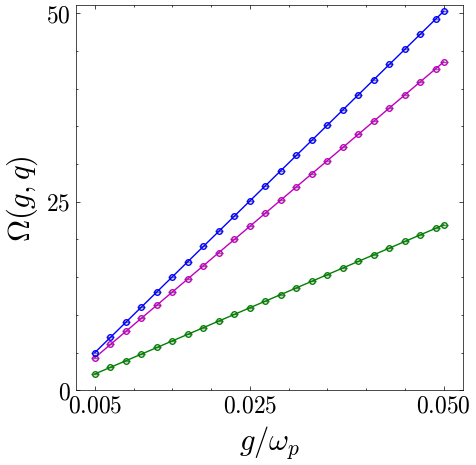}
    \caption{}
\end{subfigure}   
\hspace{0.1cm}
\begin{subfigure}[b]{0.48\textwidth}
    \centering
    \includegraphics[width=\textwidth]{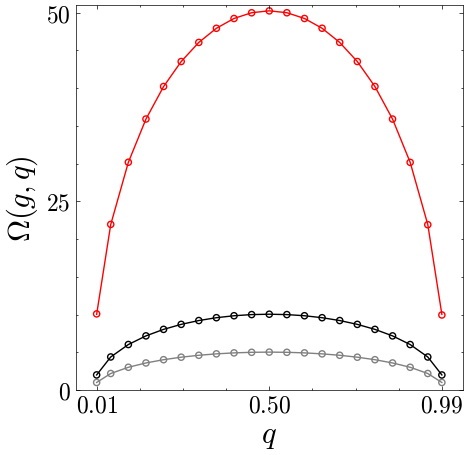}
    \caption{}
\end{subfigure} 
\caption{(a): Frequency $\Omega(g,q)$ (in units of the scale $\tilde{\Omega}\equiv\Omega(g=0.005\omega_p,q=0.01)$) as a function of $g/\omega_p$ for different values of $q$: 0.05 (green), 0.25 (purple) and 0.5 (blue). (b): $\Omega(g,q)$ (in units of the scale $\tilde{\Omega}\equiv\Omega(g=0.005\omega_p,q=0.01)$) as a function of $q$ for different values of $g/\omega_p$: 0.005 (grey), 0.01 (black) and 0.05 (red). Circles represent the numerical values obtained from the stored energy curves, solid lines represent the fit function.
Other parameters are: $E_J/E_C=100$, $\tau=\tau_p$ and $c=1$ }
\label{fig:FREQ FIT}
\end{figure}

The aim of this section is to characterize the behavior of the stored energy in a more detailed way. In lack of an analytical expression for the dynamics of the transmon under collisional charging, its behavior as a function of $n,g,q$ can be guessed phenomenologically from the damped oscillations shown so far, leading to the expression
\begin{equation}
    \Delta E(n,g,q)=\frac{E_f}{2}\bigg[1-e^{-\Gamma(g,q)n}\cos\bigl(\Omega(g,q)n\bigr)\bigg]
    \label{eq:fit_func}
\end{equation}
with $\Omega(g,q)$ the frequency of its oscillations and $\Gamma(g,q)$ the damping coefficient. 
Results for these parameters as functions of $g$ and $q$ have been obtained by fitting the stored energy curves with Eq.~(\ref{eq:fit_func}), for many different values of $q$ and $g$.\\
Starting from the frequency, we are going to prove the ansatz in Eq.~(\ref{eq:ansatz_omega}), focusing in particular on the first oscillations, where we can neglect the damping. These oscillations also are the most relevant ones in view of applications, where fast charging is important.
Results for $\Omega(g,q)$ are shown in Fig.~\ref{fig:FREQ FIT}.\\
\\
As shown in Fig.~\ref{fig:FREQ FIT}(a), the frequency grows linearly with the coupling $g/\omega_p$ regardless of the value of $q$, which only determines the slope. Also, as first noticed in the previous section, at a given coupling the frequency is maximal when $q=0.5$. This can be seen in Fig.~\ref{fig:FREQ FIT}(b) where we considered the frequency as a function of $q$ for different values of $g/\omega_p$. As expected from the density plot of the stored energy, $\Omega(g,q)$ is symmetric with respect to the $q=0.5$ axis.
Fitting its values with the ansatz
\begin{equation}
    \label{eq:Frequency(g,q)}
    \Omega(g,q) = \Omega^* \bigg(\frac{g}{\omega_p}\bigg)^\alpha [q(1-q)]^\beta 
\end{equation}
we find that the best values of the above parameters are $\alpha\approx1$, $\beta\approx\frac{1}{2}$ and $\Omega^*\approx1.7$, so that the $q$-dependent term in Eq.~(\ref{eq:Frequency(g,q)}) is the same factor appearing in the coherences of the chargers. Thus, the results shown above for the frequency $\Omega(g,q)$ provide further evidence that the quantum coherences of the chargers are directly responsible for the oscillatory behavior of the collisional charging.\\
\begin{figure}[h!]
\centering
\begin{subfigure}[b]{0.4675\textwidth}
    \centering
    \includegraphics[width=\textwidth]{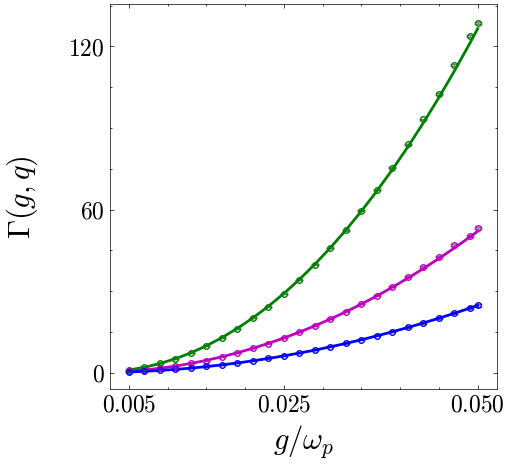}
    \caption{}
\end{subfigure}   
\hspace{0.1cm}
\begin{subfigure}[b]{0.4925\textwidth}
    \centering
    \includegraphics[width=\textwidth]{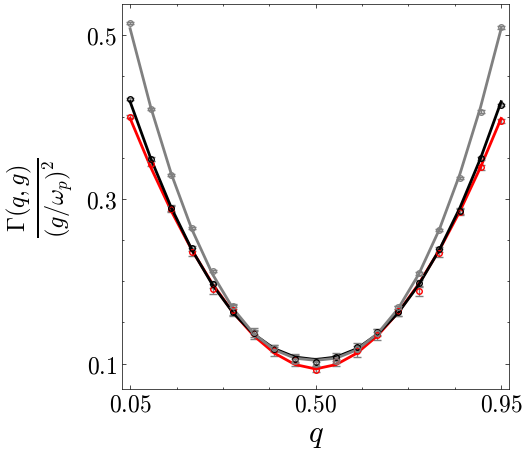}
    \caption{}
\end{subfigure} 
\caption{(a): $\Gamma(g,q)$ in units of the scale $\tilde{\Gamma}\equiv\Gamma(g=0.005\omega_p,q=0.05)$ as a function of $g/\omega_p$ for different values of $q$: 0.05 (green), 0.25 (purple) and 0.5 (blue). (b): $\frac{\Gamma(g,q)}{(g/\omega_p)^2}$ in units of the scale $\tilde{\Gamma})$ as a function of $q$ for different values of $g/\omega_p$: 0.005 (grey), 0.01 (black) and 0.05 (red). Circles represent the numerical values obtained from the stored energy curves, solid lines represent the fit function.
Other parameters are: $E_J/E_C=100$, $\tau=\tau_p$ and $c=1$ }
\label{fig:DAMPING FIT}
\end{figure}

Focusing now on the damping coefficient, results are shown in Fig.~\ref{fig:DAMPING FIT}. As hinted in Sec.~\ref{sec:sec_coherent_weak}  the damping effects become more relevant at higher couplings. It is also symmetric with respect to $q=0.5$, where it reaches its minimal value, for all the couplings considered. This is in agreement with the previous discussion and represents a very relevant feature: maintaining maximum quantum coherences at the level of the chargers protects the stored energy oscillations from the damping.
Fitting its values with the ansatz
\begin{equation}
    \label{eq:Damping(g,q)}
    \Gamma(g,q) = \Gamma^* \bigg(\frac{g}{\omega_p}\bigg)^\beta \bigg[\bigg(q-\frac{1}{2}\bigg)^\delta+c\bigg] 
\end{equation}
we find that $\beta\approx2$ for each of the curves in Fig.~\ref{fig:DAMPING FIT}(a). For what concerns the other parameters, their values show a residual slight dependence on the coupling which is summarized in Table~\ref{tab:damping_parameters} for the curves shown in Fig.~\ref{fig:DAMPING FIT}(b). \\
\begin{table}[h!]
    \centering
    \begin{tabular}{|c|c|c|c|}
    \hline
    \textbf{$g/\omega_p$}	& \textbf{$\Gamma^*$}	& \textbf{$\delta$} & \textbf{$c$}\\
    \hline
    $5\times 10^{-3}$ & 1.32 & 1.84 & 0.070\\
    \hline
    $1\times 10^{-2}$ & 1.69 & 2.11 & 0.062\\
    \hline
    $5\times 10^{-2}$ & 2.52 & 2.29 & 0.041\\
    \hline
    \end{tabular}
    \caption{Values of the fit parameters for $\Gamma(g,q)$ for the three coupling values shown in Fig.~\ref{fig:DAMPING FIT}(b).}
    \label{tab:damping_parameters}
\end{table}

\subsubsection{Different durations of the interaction.}
\begin{figure}[h!]
\centering
\begin{subfigure}[b]{0.48\textwidth}
    \centering
    \includegraphics[width=\textwidth]{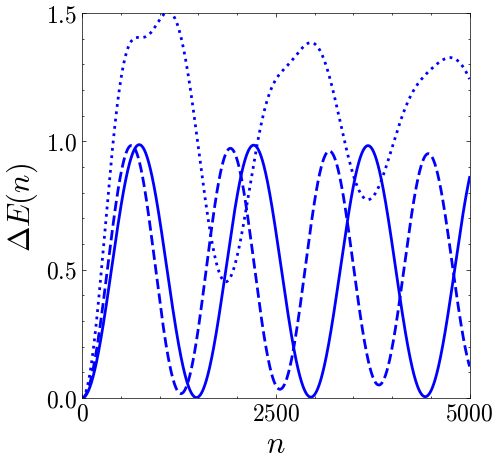}
    \caption{}
\end{subfigure}   
\hspace{0.1cm}
\begin{subfigure}[b]{0.48\textwidth}
    \centering
    \includegraphics[width=\textwidth]{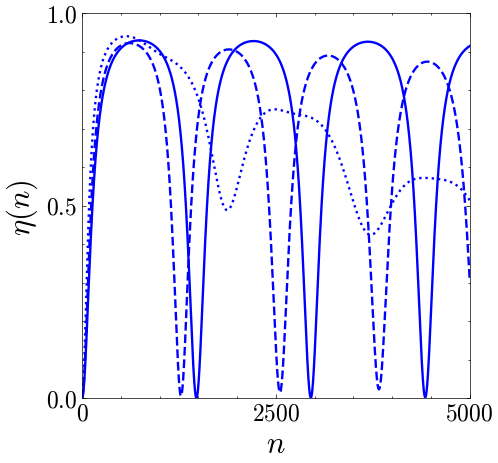}
    \caption{}
\end{subfigure} 
\caption{Stored energy $\Delta E(n)$ (in units of $E_{f}$) (a) and extraction efficiency $\eta(n)$ (b) of the QB as functions of the number of collisions $n$ for different values of $\tau = \tau_p$ (solid line), $\tau\approx1.98\tau_p $ (dashed line), $\tau\approx2.83\tau_p$ (dotted line). Other parameters are: $E_J/E_C=100$, $g/\omega_p=0.005$, $q=0.5$ and $c=1$.}
\label{fig:INCREASED_TAU_WEAK_COUPL}
\end{figure}

\begin{figure}[h!]
    \centering
    \includegraphics[width=0.5\textwidth]{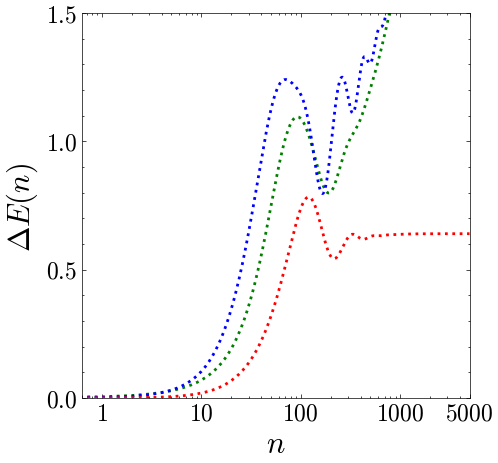}
    \caption{Stored energy $\Delta E(n)$ (in units of $E_{f}$) as a function of the number of collisions $n$ for different values of $q$: $q=0.05$ (green line), $q=0.5$ (blue line) and
    $q=0.95$ (red line)
    Other parameters are: $E_J/E_C=100$, $g/\omega_p=0.05$, $\tau\approx 2.83\tau_p$ and $c=1$.}
\label{fig:Explosive_regime}
\end{figure}

After having considered the consequences of increasing the coupling, let us now turn to what happens when varying the duration $\tau$ of each collision. As anticipated, we will focus on increasing the duration (see Appendix A for further details). We will discuss the fastest charging case, namely $q=0.5$, since analogous results are found also for the other values of $q$.\\
We start by varying $\tau$ while keeping $g/\omega_p$ in the lowest part of the interval considered. Results for the energy storage and extraction are shown in Fig.~\ref{fig:INCREASED_TAU_WEAK_COUPL}.
Oscillations like the ones pointed out for $\tau=\tau_p$ are still present for the intermediate duration considered ($\tau\approx2\tau_p$). This is of great relevance since it shows that a perfect fine-tuning $\tau=\tau_p$ is not needed to store energy in the battery. Also, we see that for this larger value of the interaction time, the oscillations are a little faster.\\
However, by further increasing the duration one observes a different behavior both at the level of the stored energy (Fig.~\ref{fig:INCREASED_TAU_WEAK_COUPL} (a)) and energy extraction efficiency (Fig.~\ref{fig:INCREASED_TAU_WEAK_COUPL} (b)). In fact, focusing in particular on the former, the periodic and weakly damped oscillations are lost, replaced by irregular beats that exceed the maximum amount of energy that can be stored in the well.  
This behavior is due to the fact that the non-bound states of the transmon QB are relevantly populated, and it becomes even more irregular when increasing the coupling. This is reported in Fig.~\ref{fig:Explosive_regime}, which also shows that the symmetry of the stored energy with respect to $q \rightarrow 1-q$ is lost in this regime. This is a signature of the fact that, in this regime, the coherences of the ancillas are not the only responsible of the charging dynamics.\\
The rapidly increasing stored energy in this long-duration regime can be useful if one is interested in storing as much energy as possible.  However, oscillations like the ones obtained for smaller durations guarantee a long-times repeatability, which can be useful to maintain the control over the system and its stability.\\

All along this section we have used the parameter $q$ to control both coherences and populations. Deviations from the maximally coherent situation at fixed $q$ can be examined by reducing the $c$ parameter below $c=1$. This takes into account the possibility of decoherence at the level of the ancillas waiting for the interaction with the QB. Appendix B is devoted to a discussion of this case.


\subsection{Incoherent charging}
\label{sec:sec_incoherent}
In this Section we compare the coherent and incoherent charging protocols. In the incoherent case ($c=0$) the chargers can only contribute via their populations, controlled by the parameter $q$
which goes from $q=1$ (ground state) to $q=0$ (excited state).

\subsubsection{Collision time $\tau=\tau_p$}
\begin{figure}[h!]
\centering
\begin{subfigure}[b]{0.48\textwidth}
    \centering
    \includegraphics[width=\textwidth]{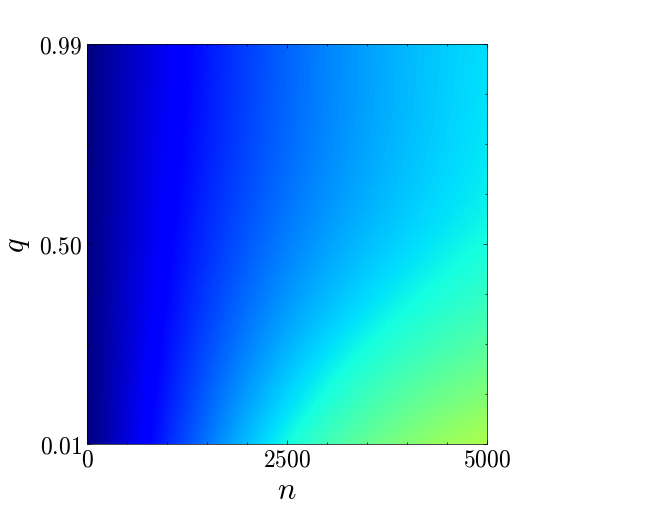}
    \caption{}
\end{subfigure}   
\hspace{0.1cm}
\begin{subfigure}[b]{0.48\textwidth}
    \centering
    \includegraphics[width=\textwidth]{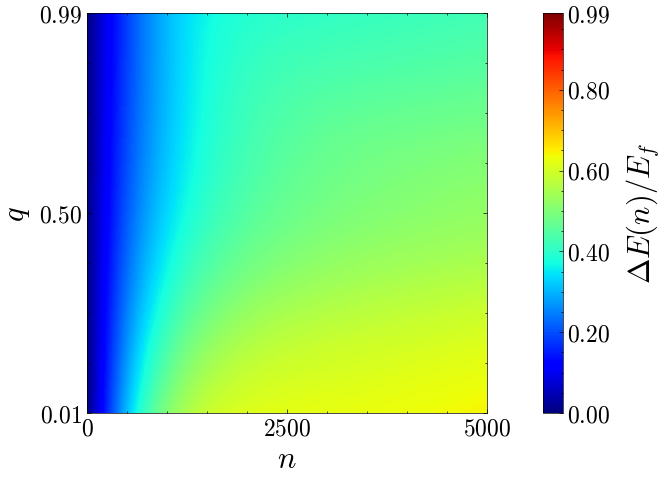}
    \caption{}
\end{subfigure} 
\caption{Density plots of the stored energy $\Delta E(n,q)$ (units of $E_{f}$) as a function of the number of collisions $n$ and of the ancillary parameter $q$ for: (a) $g=2.5\times 10^{-2}\omega_p$, (b) $g=5\times 10^{-2}\omega_p$. Other parameters are: $E_J/E_C=100$, $\tau=\tau_p$ and $c=0$.}
\label{fig:EN TAU_p INCOER}
\end{figure}

According to the analysis carried out in Sec. \ref{sec:sec_coherent}, it clearly emerges that the charging dynamics crucially depends on the coherences of the chargers. Therefore, by turning them off, we expect the stored energy to be negligible with respect to the coherent case. In fact, this is the case. Moreover, we needed to move towards greater values of the coupling in order to obtain an incoherent stored energy comparable to the coherent one. Numerical results reported in Fig.~\ref{fig:EN TAU_p INCOER} show the density plot of the stored energy as a function of both $n$ and $q$ for two representative values of the coupling. 

\begin{figure}[h!]
\centering
\begin{subfigure}[b]{0.48\textwidth}
    \centering
    \includegraphics[width=\textwidth]{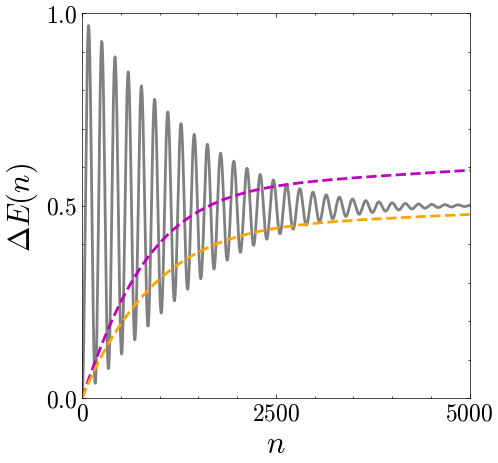}
    \caption{}
\end{subfigure}   
\hspace{0.1cm}
\begin{subfigure}[b]{0.48\textwidth}
    \centering
    \includegraphics[width=\textwidth]{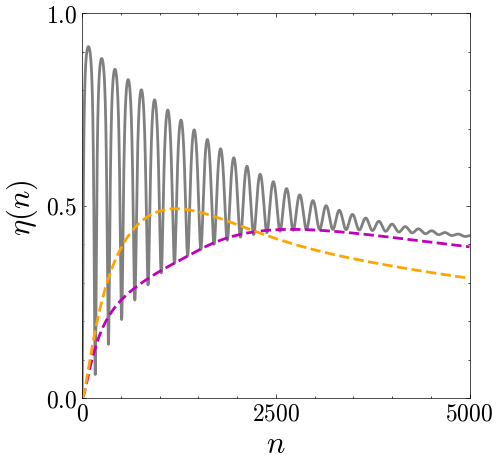}
    \caption{}
\end{subfigure} 
\caption{Stored energy $\Delta E(n)$ (in units of $E_{f}$) (a) and extraction efficiency $\eta(n)$ (b) as functions of the number of collisions $n$. Comparison between the coherent protocol for $q=0.25$ and $q=0.75$ (solid grey line) with the corresponding incoherent ones: dashed purple for $q=0.25$; dashed orange for $q=0.75$. Other parameters are: $E_J/E_C=100$, $\tau=\tau_p$, $g=5\times10^{-2}\omega_p$.}
\label{fig:TAUp_STRONGER_COUPL_COER_VS_INCOER}
\end{figure}

\begin{figure}[h!]
\centering
\begin{subfigure}[b]{0.48\textwidth}
    \centering
    \includegraphics[width=\textwidth]{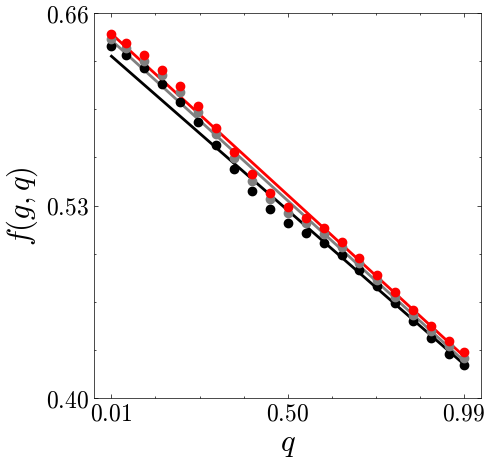}
    \caption{}
\end{subfigure}   
\hspace{0.1cm}
\begin{subfigure}[b]{0.48\textwidth}
    \centering
    \includegraphics[width=\textwidth]{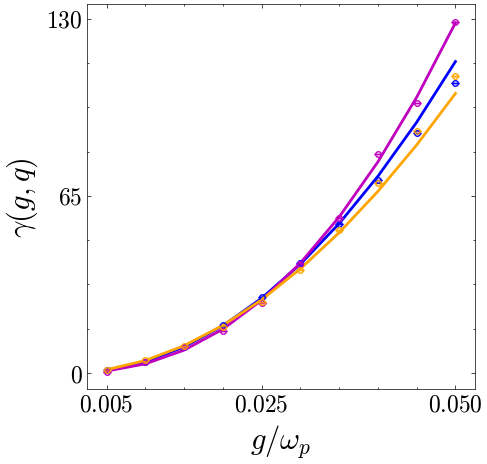}
    \caption{}
\end{subfigure} 
\caption{(a): $f(g,q)$ in units of $E_f$ as a function of $q$ for different values of $g/\omega_p$: $4\times 10^{-2}$ (grey), $5\times 10^{-2}$ (black), $6\times 10^{-2}$ (red). (b): $\gamma(g,q)$ in units of the scale $\tilde{\gamma}=\gamma(g=5\times 10^{-3}\omega_p,q=0.5)$ as a function of $g/\omega_p$ for different values of $q$: 0.5 (blue), 0.25 (purple) and 0.75 (orange). Circles represent the numerical values obtained from the stored energy curves, solid lines represent the fit function. Other parameters are: $E_J/E_C=100$, $\tau=\tau_p$ and $c=0$.}
\label{fig:INCOER_FIT}
\end{figure}

First of all we notice that the oscillating behavior of $\Delta E(n)$ disappears. However, the greater coupling allows the storage of a significant amount of energy which, at long times, becomes comparable with the coherent case stored energy and which is no more invariant under $q \rightarrow 1-q$. As expected in this case, the charging performances are better near $q=0$, namely for (almost) full chargers.\\
The loss of the $q\rightarrow1-q$ symmetry in the incoherent case can also be visualized in Fig.~\ref{fig:TAUp_STRONGER_COUPL_COER_VS_INCOER}(a), where the coherent charging protocols with $q=0.25$ and $q=0.75$ (which led to the same $\Delta E(n)$ in the coherent case, due to the $q\rightarrow1-q$ symmetry) are compared with the respective incoherent protocols.
Fig.~\ref{fig:TAUp_STRONGER_COUPL_COER_VS_INCOER} highlights other remarkable differences:
\begin{itemize}
    \item at short times the energy stored in the coherent case is greater than the one of the incoherent case;
    \item at long times the coherent oscillations are damped until reaching a value which is comparable with the stored energy of the incoherent case and, in some cases, even lower.
\end{itemize}
Fig.~\ref{fig:TAUp_STRONGER_COUPL_COER_VS_INCOER} (b) shows the corresponding results for the extraction efficiency. In the incoherent case, it barely reaches $\eta=50 \%$. Thus, provided being able to stop the evolution at short times, the coherent charging protocol reveals to be convenient also for higher values of the coupling.\\
\\
Let us now turn to a more detailed study of the stored energy of the incoherent protocol. Given previous results, we assume it to be of the form
\begin{equation}
\label{eq:fit_func_incoer}
    \Delta E(n,g,q)=f(g,q)\bigg[1-e^{-\gamma(g,q)n}\bigg],
\end{equation}
with $\gamma(g,q)$ representing the inverse of the charging time of the battery and $f(g,q)$ the asymptotic value reached after the stored energy flattens. Numerical values for these parameters have been obtained by fitting the stored energy curves with Eq.~(\ref{eq:fit_func_incoer}) for different values of $g$ and $q$. The results for the fit procedures are shown in Fig.~\ref{fig:INCOER_FIT}.\\
Fig.~\ref{fig:INCOER_FIT}(a) shows that the asymptotic value $f(g,q)$ reached by the stored energy at long times is almost insensitive to coupling variations, whereas it decreases roughly linearly. However, even if $q \approx 1$, a considerable amount of energy is stored. This suggests the assumption 
\begin{equation}
    f(q) = E_f[a(1-q)+b].
\end{equation}
By fitting the numerical values obtained for the stored energy fit, one obtains $a\approx0.21$ and $b\approx 0.43$.
The origin of $b$, namely the value of the energy asymptotically stored in the battery when $q\rightarrow 1$, can be traced back to the energetic contribution needed to switch-on and off each QB-charger coupling~\cite{Barra19, Barra22}: this splits in two parts, one is gained by the ancillas whereas one is transferred directly to the QB. Such switching contributions arise whenever the battery and interaction Hamiltonians do not commute with each other~\cite{Andolina18}, as happens in the present case.\\
\\
In contrast, Fig.~\ref{fig:INCOER_FIT}(b) shows 
a completely different behavior for $\gamma(g,q)$, which represents the inverse of the charging time scale of the incoherent protocol: it strongly increases with $g/\omega_p$, whereas it is almost insensitive to variations of $q$. Assuming consistently that 
\begin{equation}
    \gamma(g,q) = \gamma^*(q) \bigg(\frac{g}{\omega_p}\bigg)^\beta
\end{equation}
the optimal values for the parameter $\beta$ to fit the numerical data obtained is $\beta \approx2$, whereas $\gamma^*$ is a scale of the order of $10^{-1}$ with a slight dependence from $q$ which, however, is negligible with respect to the dependence on $g$. \\
This is in agreement with what has been anticipated at the beginning of this Section: the incoherent protocol at weaker couplings is characterized by a small $\gamma$, which corresponds to a large charging time, so that one needs a huge amount of collisions to store a significant amount of energy into the QB.\\

\subsubsection{Different durations of the interactions}
\begin{figure}[h!]
\centering
\begin{subfigure}[b]{0.48\textwidth}
    \centering
    \includegraphics[width=\textwidth]{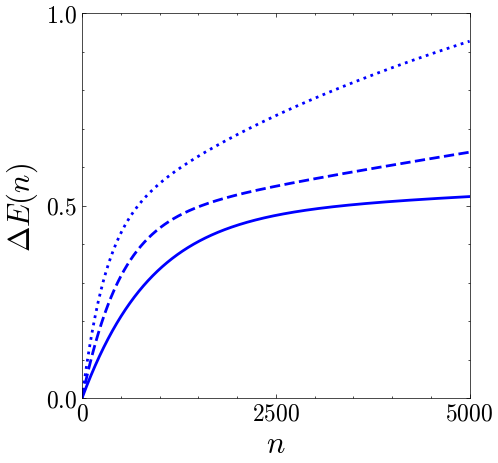}
    \caption{}
\end{subfigure}   
\hspace{0.1cm}
\begin{subfigure}[b]{0.48\textwidth}
    \centering
    \includegraphics[width=\textwidth]{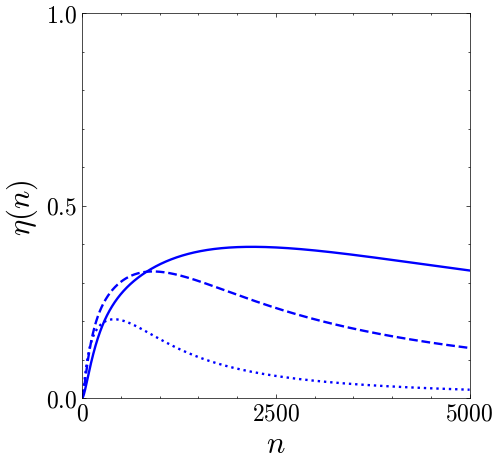}
    \caption{}
\end{subfigure} 
\caption{Stored energy $\Delta E(n)$ (in units of $E_{f}$) (a) and extraction efficiency $\eta(n)$ (b) of the QB as functions of the number of collisions $n$ for different values of $\tau = \tau_p$ (solid), $\tau\approx1.98\tau_p $ (dashed), $\tau\approx2.83\tau_p$ (dotted). Other parameters are: $E_J/E_C=100$, $g/\omega_p=0.05$, $q=0.5$ and $c=0$.}
\label{fig:INCOER_LONG_TAU}
\end{figure}

Also in the incoherent case, variations of the duration $\tau$ can affect the dependence of the stored energy on the number of collisions. This clearly emerges from Fig.~\ref{fig:INCOER_LONG_TAU}, where we compare incoherent charging protocols with different values of $\tau$.
\\
The interesting feature is that, for $\tau>\tau_p$, the stored energy increases without stabilizing at a given value. Similarly to the coherent case, the behavior at higher $\tau$ can be interesting in view of maximizing the stored energy. However, in the incoherent case increasing the duration has a negative effect on the extraction efficiency $\eta(n)$. Considering for example $\tau\approx 2.83 \tau_p$ we see that, even if the stored energy after $5000$ collisions is almost doubled with respect to $\tau=\tau_p$, $\eta(n)$ has strongly reduced, so that there is no convenience in moving towards such high values of the collision time.

\subsection{Experimental feasibility}
\label{sec:Exp_feas}
\begin{figure}[h]
    \centering
    \begin{circuitikz}
        \ctikzset{capacitors/scale=0.7}
        \draw
        (-4.5,0.5) -- (-4.5,1.5)
        (-4.5,0.5) -- (-3.5,0.5)
        (-3.5,0.5) -- (-3.5,1.5)
        (-4.5,1.5) -- (-3.5,1.5)
        (-4.5,1.5) -- (-3.5,0.5)
        (-4.5,0.5) -- (-3.5,1.5)
        (-2.5,2.5) -- (-2.5,3)
        (-4,2.5) -- (-4,1.5)
        (-4,-0.5) -- (-4,0.5)
        (-1,2.5) to[C, l^=$C_B$,name=Cb] (-1,-0.5)
        (-4,-0.5) -- (-1,-0.5)
        (-4,2.5) -- (-1,2.5)
        (-2.5,-0.5) -- (-2.5,-1)
        
        (2,0.5) -- (2,1.5)
        (2,0.5) -- (3,0.5)
        (3,0.5) -- (3,1.5)
        (2,1.5) -- (3,1.5)
        (2,1.5) -- (3,0.5)
        (2,0.5) -- (3,1.5)
        (2.5,1.5) -- (2.5, 2.5)
        (2.5,2.5) -- (5.5,2.5)
        (2.5,0.5) -- (2.5,-0.5)
        (5.5,2.5) to[C, l^=$C_n$,name=Cb] (5.5,-0.5)
        (2.5,-0.5) -- (5.5, -0.5)
        (4,3) to[C, l^=$C_{Bn}$,name=Cg] (-2.5,3)
        (4,3) -- (4,2.5)
        (4,-0.5) -- (4, -1)

        (-3,-1) -- (-2,-1)
        (-2.9,-1.1) -- (-2.1,-1.1)
        (-2.8,-1.2) -- (-2.2,-1.2)
        (-2.6,-1.3) -- (-2.4,-1.3)

        (3.5,-1) -- (4.5,-1)
        (3.6,-1.1) -- (4.4,-1.1)
        (3.7,-1.2) -- (4.3,-1.2)
        (3.9,-1.3) -- (4.1,-1.3)
        ;
    \end{circuitikz}
    \vspace*{1cm}
    \caption{Scheme of two capacitively coupled transmon circuits. $C_B$ and $C_n$ represent the shunting capacitance of the transmon QB and of the transmon ancilla respectively, whereas $C_{Bn}$ is the capacitance employed to couple them. The SQUID of each transmon has been denoted with the crossed boxes.}
    \label{fig:TRANSMON_coupling}
\end{figure}
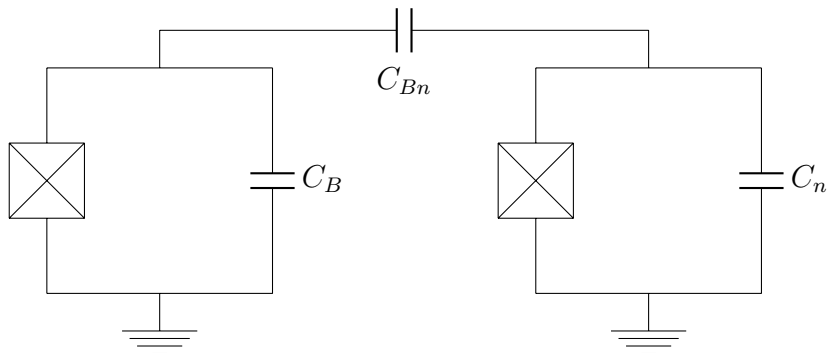

We conclude this Section by addressing the features which real superconducting platforms must possess to implement the QB described in this work.\\
As stated above, the anharmonic multilevel QB can be realized through the superconducting circuit in Fig.~\ref{fig:TRANSMON}. In particular, the transmon regime $E_J\gg E_C$ guarantees an anharmonic and charge insensitive multilevel structure which is suitable for our purposes. The interval usually identified as optimal for the transmon circuits is $10 \lesssim E_J/E_C \lesssim 10^4$ \cite{Koch}. Here, we considered $E_J/E_C=100$, a value that perfectly fits these bounds.\\
Assuming this and requiring $\omega_p$ to reach the typical operational frequency for a transmon $\omega_p\approx 1$ GHz, \cite{Koch, Kockum2019,Blais21} forces $E_C$ to be of the order of $E_C \approx 30$ MHz. Considering again Fig.~\ref{fig:TRANSMON}, it is straightforward to show that, if the shunting capacitance $C_B$ is large enough, one has \cite{Koch} 
\begin{equation}
    E_C = \frac{e^2}{2 C_{tot}} \approx \frac{e^2}{2C_B},
\end{equation}
since $C_{tot}=C_B+C_J+C_g\approx C_B$.
This implies that the required order of magnitude of $E_C$ is achieved by assuming $C_B \approx 3$ pF.
This value fits the typical range of capacitances used for superconducting circuits which, due to their typical size, is often in the fF to low pF range \cite{Roth_2023,Blais21}. \\
For what concerns one single ancillary qubit, it can be realized through a transmon circuit with a greater anharmonicity $\alpha_r'$ so that, by initializing it as a superposition of its ground and first excited state, one can neglect its multilevel structure and describe it as an effective TLS realized addressing the two lower energetic states. In order to increase its anharmonicity one could choose values of $E_J'/E_C'$ near to the lower bound of the transmon range, namely $E_J'/E_C' \approx 10$, a condition such that $\alpha_r'\approx3\alpha_r$.\\
The capacitive coupling between two transmon circuits can be realized through the circuit represented in Fig.~\ref{fig:TRANSMON_coupling} and further controlled in time implementing quantum couplers~\cite{Sete21, Campbell23, Heunisch23}. By doing so, the coupling Hamiltonian takes the form~\cite{Krantz_2019}
\begin{equation}
    \hat{V}_{B,n}= \tilde{g}\hat{N} \hat{N}_n
\end{equation}
with 
\begin{equation}
    \tilde{g}=\frac{4 e^2 C_{Bn}}{C_B C_n}.
\end{equation}
Within the TLS assumption for the ancillary transmon circuit, we can project $\hat{N}_n$ in the basis $\{|0\rangle_n,|1\rangle_n\}$, where $|0\rangle_n$ and $|1\rangle_n$ respectively represent the ground and first excited states of the ancilla, obtaining
\begin{equation}
    \hat{N}_n = N_{1,1} |1\rangle\langle1|_n +
    N_{0,0} |0\rangle\langle0|_n + N_{1,0}( \hat{\sigma}^{-}_{n} + \hat{\sigma}^{+}_{n}).
\end{equation}
Direct numerical calculations of these coefficients show that the diagonal terms are negligible when compared with the off-diagonal terms, so that one can safely assume 
\begin{equation}
    \hat{V}_{B,n}\approx \tilde{g}N_{1,0} \hat{N} (\hat{\sigma}^{-}_{n} + \hat{\sigma}^{+}_{n})\equiv g \hat{N} (\hat{\sigma}^{-}_{n} + \hat{\sigma}^{+}_{n}).
\end{equation}
Thus, the resulting ratio between the coupling and the plasma frequency takes the form
\begin{equation}
    \frac{g}{\omega_p} = 4 e^2 N_{1,0} \frac{C_{Bn}}{C_B C_n} \frac{1}{\sqrt{8E_J E_C}}.
\end{equation}
Assuming $E_J=100E_C$ one obtains
\begin{equation}
    \frac{g}{\omega_p} = \frac{4 e^2 N_{1,0}}{10\sqrt{8}} \frac{C_{Bn}}{C_B C_n} \frac{1}{E_C}=\frac{\sqrt{2}N_{1,0}}{5}\frac{C_{Bn}}{C_n}.
\end{equation}
Considering that $\frac{\sqrt{2}N_{1,0}}{5} \approx 0.4$, in order to work in the coupling regime explored in \ref{sec:sec_coherent_weak}, one needs $1\times 10^{-2}\lesssim C_{Bn}/C_n \lesssim 1\times10^{-1}$. Assuming $C_n\approx C_B \approx$ pF, the required coupling capacitance reaches values of $C_{Bn} \approx 10-100$ fF, which again fit the typical range for the capacitance of superconducting circuits mentioned above.\\
With the parameters discussed above, the duration $\tau_p=1/\omega_p$ is of the order of a few nanoseconds. This value fits the standard manipulation times for superconducting circuits like transmons \cite{Koch}.\\
The realization of the sequential interaction between the transmon QB and a set of uncorrelated and non interacting ancillary transmons can be implemented in two ways. The first and more direct one is to realize many identical ancillary transmons, to keep them isolated from each other, and to independently and sequentially couple them with the transmon battery in a configuration analogous to a very highly coordinated quantum computers.\\
A problem which may arise in this case is the survival of quantum coherences on a set of qubits large enough to obtain a significant energy storage ($400-1000$ collisions as previously seen).
Indeed, the global charging time $t_c$ of the battery can be estimated at most as $t_c \approx 1\mu s$, corresponding to $1000$ interaction of the order of nanoseconds. This time scale competes with the decoherence and dephasing times $T_1$ and $T_2$ of transmon circuits, which represent the time scales over which quantum coherences are suppressed by the system environment interaction and quantum phases are randomized. However, usual parameters for transmons circuits lead to  $T_1,T_2 \gtrsim 10t_c$ \cite{Koch}, so that the the coherence of trasmon ancillas can be maintained for more than one oscillation period without relevant losses. \\ 
The main problem of this direct approach is the possibility of realizing such a great number of qubits and to keep them isolated from each other. For this reason, one could be interested in exploring the possibility of realizing only a small number $N'$ of ancillary qubits, each of which could be re-initialized to the same initial state after the interaction with the QB, in order to be reused after the battery has interacted with all of the $N'$ ancillas. This would be even better for what concerns the maintainace of ancillary quantum coherences, since they would only need to survive for a smaller time $t_r\approx N'\tau$. 
In principle, this scheme could be brought to an extreme considering only one ancilla ($N'=1$). This option, however, is not optimal, since during the time needed for the re-initialization the charging stops. The realization of multiple ancillary qubits would guarantee the continuous charging, while the used ancillas are progressively re-initialized. \\
Such a reuse protocol for ancillary qubits has been proposed very recently in~\cite{muniz25} in a different quantum computing platform based on neutral atoms. The technical challenge would therefore be represented by the adaptation of this re-initialization process to superconducting case.\\


\section{Conclusions}
\label{Conclusions}
In this work we have characterized the charging and energy extraction efficiency of a quantum battery based on a superconducting quantum circuit in the transmon regime~\cite{Dou23}. Through numerical analysis, we have shown that the charging realized by means of the sequential interaction of the quantum battery with independent and coherent ancillas leads to a (slightly damped) oscillating behavior as a function of the number of collisions. The functional dependence of the energy stored into the device on the battery-ancilla coupling and the coherence within the ancillas has been reconstructed through best fits. For short enough duration of the collisions this energy remains stably confined within the cosine potential well given by the Josephson energy of the transmon and, in correspondence with the first maxima of the oscillations, the energy stored into the device can be almost completely extracted as useful work. For longer times of interaction, the stored energy is greater but less stable. Moreover, it cannot be completely extracted. When the charging is realized by means of incoherent ancillas the oscillations disappear and the asymptotic value of the stored energy reached after a great number of collisions depends on the coupling and on the energy initially trapped into each ancilla. \\
The major differences between these two regimes could be understood starting from mean field considerations about the effective Hamiltonian experienced by the transmon. Indeed, if one considers the effective contribution coming from an averaging of the $\hat{H}_{SB}$ Hamiltonian over the ancillary state $\hat{\eta}_n$ at the beginning of each collision, it is non-zero only in the presence of coherences ($c\neq0$). Thus, in the absence of quantum coherences higher-order moments must intervene to enhance the energy transfer to the battery, resulting in a slower dynamics. A deeper understanding could be provided by recalling results already discussed in literature~\cite{Seah} for a harmonic oscillator quantum battery charged through a collisional charging scheme. Here, the discrete dynamics of the incoherent charging has been mapped into a classical discrete time random walk, with the interference effects of quantum coherences leading to a quantum random walk which is ultimately responsible for the speed-up of the battery charging.

The proposed scheme for a transmon quantum battery combines the possibility to reach a high level of control on the energy storage to a great extraction efficiency. Moreover, its working point is in a regime of parameter within reach in nowadays superconducting circuits. According to this, the present work can contribute to provide a robust theoretical background for future experimental implementations of quantum batteries in state of the art solid state platforms.



\paragraph{\bf{Acknowledgements}}
Authors would like to thank G. Benenti and M. Sassetti for useful discussions. We also thank anonymous Referees for their useful comments and suggestions.\\
D. F. and N. M. acknowledge the contribution of the European Union-NextGenerationEU through the "Quantum Busses for Coherent Energy Transfer" (QUBERT) project, in the framework of the Curiosity-Driven 2021 initiative of the University of Genova. F. C and D. F. acknowledge the contribution of the European Union-NextGenerationEU through the "Solid State Quantum Batteries: Characterization and Optimization" (SoS-QuBa) project (Prot. 2022XK5CPX), in the framework of the PRIN 2022 initiative of the Italian Ministry
of University (MUR) for the National Research Program
(PNR). This project has been funded within the programme ``PNRR Missione 4 - Componente 2 - Investimento 1.1 Fondo per il Programma Nazionale di Ricerca e Progetti di Rilevante Interesse Nazionale (PRIN)''.

\newpage
\section*{Appendix A: $\tau<\tau_p$}
\label{sec:App_A}
\subsection*{A.1: Coherent charging}
\begin{figure}[h!]
\centering
\begin{subfigure}[b]{0.48\textwidth}
    \centering
    \includegraphics[width=\textwidth]{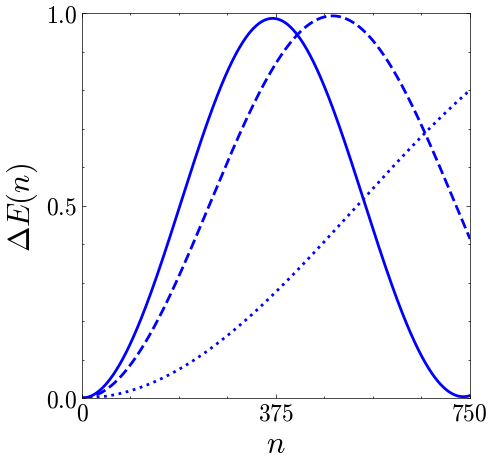}
    \caption{}
\end{subfigure}   
\hspace{0.1cm}
\begin{subfigure}[b]{0.48\textwidth}
    \centering
    \includegraphics[width=\textwidth]{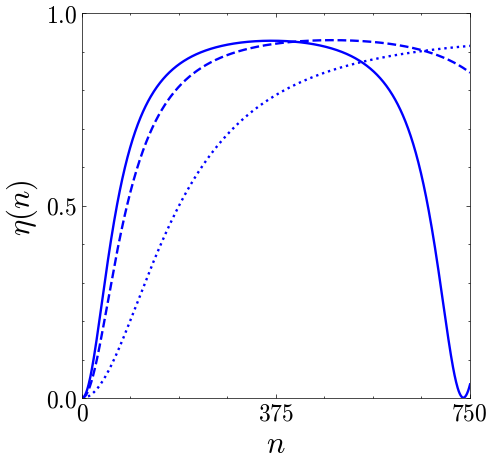}
    \caption{}
\end{subfigure} 
\caption{Stored energy $\Delta E(n)$ (in units of $E_{f}$) (a) and extraction efficiency $\eta(n)$ (b) as functions of the number of collisions $n$ for $\tau=\tau_p$ (solid), $\tau=0.7\tau_p$ (dashed), $\tau=0.3\tau_p$ (dotted). Other parameters are: $E_J/E_C=100$, $q=0.5$, $g/\omega_p=1\times10^{-2}$, $c=1$.}
\label{fig:smaller_tau}
\end{figure}

Results obtained by reducing the interaction duration $\tau$ below $\tau_p$ are shown in Fig.~\ref{fig:smaller_tau} for the coherent case and for $g=1\times10^{-2}\omega_p$.\\
\\
The oscillating dynamics of the QB, for both the storage and extraction becomes slower if the duration is decreased. Thus, in view of experimental implementations, there is no advantage in reaching the regime $\tau < \tau_p$: it is challenging at the technical level and worsens the QB charging performance in terms of charging time, since a greater amount of collisions is needed to reach the top of the well.
\subsection*{A.2: Incoherent charging}
\begin{figure}[h!]
\centering
\begin{subfigure}[b]{0.48\textwidth}
    \centering
    \includegraphics[width=\textwidth]{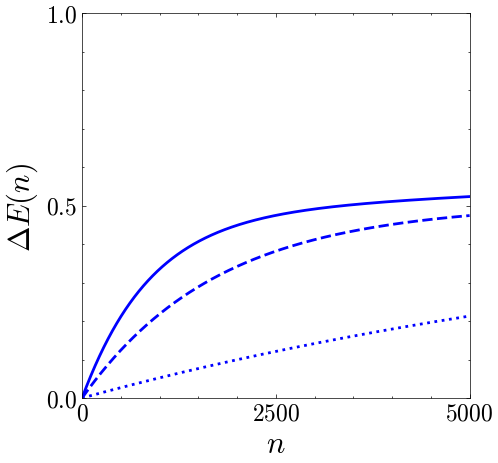}
    \caption{}
\end{subfigure}   
\hspace{0.1cm}
\begin{subfigure}[b]{0.48\textwidth}
    \centering
    \includegraphics[width=\textwidth]{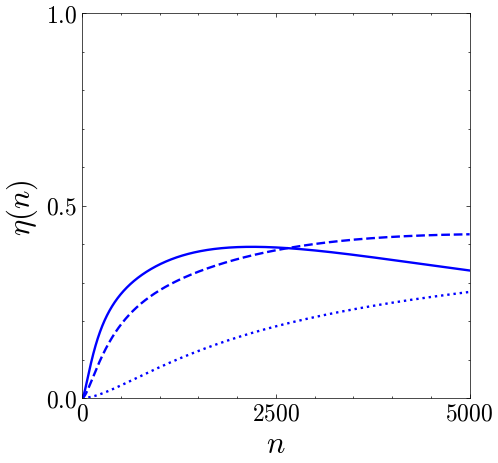}
    \caption{}
\end{subfigure} 
\caption{Stored energy $\Delta E(n)$ (in units of $E_{f}$) (a) and extraction efficiency $\eta(n)$ (b) as functions of the number of collisions $n$ for $\tau=\tau_p$ (solid), $\tau=0.7\tau_p$ (dashed), $\tau=0.3\tau_p$ (dotted). Other parameters are: $E_J/E_C=100$, $q=0.5$, $g/\omega_p=5\times10^{-2}$, $c=1$.}
\label{fig:smaller_tau_inc}
\end{figure}

For what concerns the incoherent case, the results obtained for $\tau<\tau_p$ are shown in Fig.~\ref{fig:smaller_tau_inc}. As in the coherent case, the charging is slower at smaller values of $\tau$. One interesting fact is that the long time value of efficiency $\eta(n)$ for $\tau=0.7\tau_p$ is higher than the corresponding value for $\tau=\tau_p$. Since the long time stored energy are comparable, one could consider, despite the technical challenges, to slightly reducing the duration of the collisions to enhance the extraction performances.
\\

\section*{Appendix B: Considerations about intermediate of $c$}
\label{sec:App_B}
\begin{figure}[h!]
\centering
\begin{subfigure}[b]{0.48\textwidth}
    \centering
    \includegraphics[width=\textwidth]{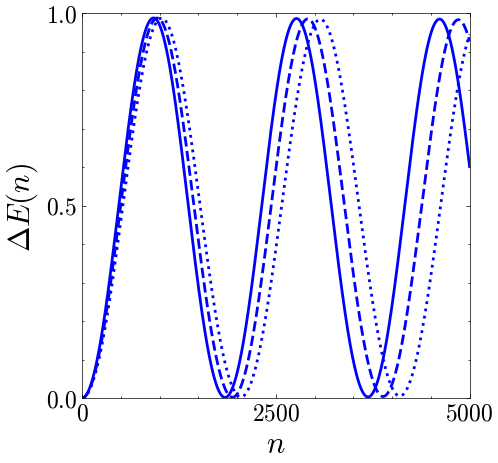}
    \caption{}
\end{subfigure}   
\hspace{0.1cm}
\begin{subfigure}[b]{0.48\textwidth}
    \centering
    \includegraphics[width=\textwidth]{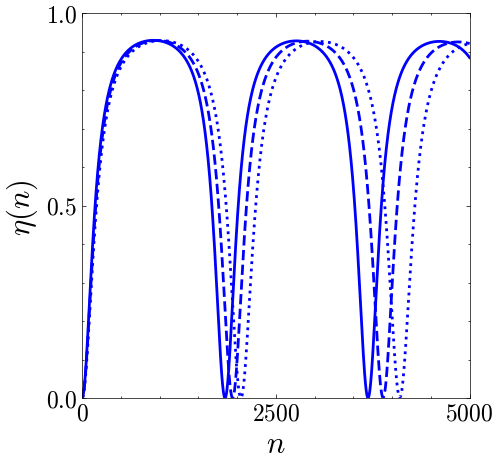}
    \caption{}
\end{subfigure} 
\caption{Stored energy $\Delta E(n)$ (in units of $E_{f}$) (a) and extraction efficiency $\eta(n)$ (b) as functions of the number of collisions $n$ for $c=1$ (solid), $c=0.95$ (dashed), $c=0.9$ (dotted). Other parameters are: $E_J/E_C=100$, $\tau = \tau_p$, $q=0.5$, $g/\omega_p=4\times10^{-3}$.}
\label{fig:smaller_c}
\end{figure}

Results obtained for $c<1$ case are shown in Fig.~\ref{fig:smaller_c} for $q=0.5$.\\
\\
As expected, small deviations from the $c=1$ condition cause a slight reduction in the frequency of stored energy oscillations, without disruptive effects on their maxima and on the extraction efficiency. Analogous results are found for the other values of $q$ already analyzed. \\
This is of great relevance since it implies that one can afford to lose a fraction of ancillary quantum coherences, for example due to decoherence phenomena, without compromising the battery functioning.

\newpage
\section*{References}
\bibliography{References}

\end{document}